\documentclass[twocolumn,pra,amsmath,amssymb,superscriptaddress,longbibliography,nofootinbib,floatfix]{revtex4-2}
\usepackage[utf8]{inputenc}
\usepackage{pgfplots}
\usepackage{bbold}
\usepackage{comment}
\usepackage[colorlinks=true, linkcolor=red, allbordercolors={white}]{hyperref}
\usepackage{algorithm}
\usepackage{algpseudocode}
\usepackage{ragged2e}
\usepackage{float}
\pgfplotsset{compat = newest}
\usepackage{soul}
\usepgfplotslibrary{colorbrewer}
\usepgfplotslibrary{groupplots}
\usetikzlibrary{arrows.meta}
\definecolor{color1}{HTML}{1f77b4}
\definecolor{color2}{HTML}{ff7f0e}
\definecolor{color3}{HTML}{2ca02c}
\definecolor{color4}{HTML}{d62728}
\definecolor{color5}{HTML}{9467bd}
\definecolor{color6}{HTML}{8c564b}
\definecolor{color7}{HTML}{e377c2}
\definecolor{color8}{HTML}{7f7f7f}
\definecolor{color9}{HTML}{bcbd22}
\definecolor{shadyblue}{rgb}{0.33,0.33,1}
\pgfplotsset{
  cycle list={color1\\color2\\color3\\color4\\color5\\color6\\color7\\color8\\color9\\},
}

\begin{document}

\title{Phases of Interacting Fibonacci Anyons on a Ladder at Half-Filling}

\author{Nico Kirchner}
%\email{nico.kirchner@tum.de}
\affiliation{Technical University of Munich, TUM School of Natural Sciences, Physics Department, 85748 Garching, Germany}
\affiliation{Munich Center for Quantum Science and Technology (MCQST), Schellingstr. 4, 80799 M{\"u}nchen, Germany}

\author{Roderich Moessner}
\affiliation{Max Planck Institute for the Physics of Complex Systems, Nöthnitzer Str. 38, 01187 Dresden, Germany}

\author{Frank Pollmann}
\affiliation{Technical University of Munich, TUM School of Natural Sciences, Physics Department, 85748 Garching, Germany}
\affiliation{Munich Center for Quantum Science and Technology (MCQST), Schellingstr. 4, 80799 M{\"u}nchen, Germany}

\author{Adam Gammon-Smith}
\affiliation{School of Physics and Astronomy, University of Nottingham, Nottingham, NG7 2RD, UK}
\affiliation{Centre for the Mathematics and Theoretical Physics of Quantum Non-Equilibrium Systems, University of Nottingham, Nottingham, NG7 2RD, UK}

\begin{abstract}
Two-dimensional many-body quantum systems can exhibit topological order and support collective excitations with anyonic  statistics different from the usual fermionic or bosonic ones. With the emergence of these exotic point-like particles, it is natural to ask what phases can arise in interacting many-anyon systems. To study this topic, we consider the particular case of Fibonacci anyons subject to an anyonic tight-binding model with nearest-neighbor repulsion on a two-leg ladder. Focusing on the case of half-filling, for low interaction strengths an ``anyonic'' metal is found, whereas for strong repulsion, the anyons form an insulating charge-density wave. Within the latter regime, we introduce an effective one-dimensional model up to sixth order in perturbation theory arising from anyonic superexchange processes. We numerically identify four distinct phases of the effective model, which we characterize using matrix product state methods. These include both the ferro- and antiferromagnetic golden chain phases, as well as phases with $\mathbb{Z}_2$ and incommensurate correlations.
\end{abstract}

\maketitle

\section{Introduction}

The paradigmatic Hubbard model~\cite{hubbard, hubbard2, hubbard3} describes electrons subject to local interactions and hopping processes on lattices and is one of the most important models in condensed matter physics. Despite its apparent simplicity, the Hubbard model is known to feature a rich variety of phases, including, e.g., metallic, Mott insulating, ferromagnetic and antiferromagnetic phases, and is even believed to describe high-temperature superconductivity~\cite{hubbard_phases}. A bosonic version of this model~\cite{BoseHubbard}, which can describe, e.g., ultracold atom experiments, narrow band superconductors and Josephson junction arrays~\cite{BoseHubbard1, BoseHubbard2}, has also been introduced and studied. The statistics of the underlying degrees of freedom thus play a crucial role and may be viewed as an ingredient for the different `classes' of Hubbard models.

More recently, models that are motivated as analogues to the Hubbard model or effective models arising in particular limits, such as the Heisenberg and the $t-J$ model, have been studied for anyons~\cite{PhysRevLett.126.163201, PhysRevB.108.155134, bonkhoff2024anyonicphasetransitions1d, GoldenChain, IntroFibAndGC, GoldenChainAndMG, PhysRevB.90.075129, anyonic_defects, PhysRevLett.103.070401, AnyonSpin1Chains, YangLeeAnyons, PhysRevLett.108.207201, PhysRevB.87.085106, PhysRevB.90.081111, FINCH2014299, GoldenLadder, PhysRevB.93.165128, PhysRevB.93.035124, AnyonLadderAyeni}, which are exotic quasiparticles featuring statistics beyond the bosonic and fermionic cases~\cite{Leinaas1977, PhysRevLett.48.1144, PhysRevLett.49.957}. These point-like quasiparticles can emerge in two-dimensional intrinsically topologically ordered systems such as fractional quantum Hall states~\cite{FQH1, FQH2, FQH3, FQH4, FQH6, FQH5} and quantum spin liquids~\cite{QSL0, QSL1, TQC2, QSL3, QSL4, QSL5, QSL6}, and feature exciting physical properties, such as topological protection from local perturbations, leading to the concept of fault-tolerant topological quantum computing~\cite{TQC1, TQC2, TQC3, TQC4, TQC5}.

Even apart from the future prospect of using anyons for quantum computing, it has been found that interactions in many-anyon systems can lead to novel quantum phases of matter that are realized on top of the already existing (parent) topologically ordered systems that support the anyons themselves. Examples for such phases include gapless phases described by conformal field theory~\cite{IntroFibAndGC, GoldenChain, GoldenChainAndMG, anyonic_defects, PhysRevLett.103.070401, AnyonSpin1Chains, YangLeeAnyons, PhysRevLett.108.207201, PhysRevB.87.085106, PhysRevB.90.081111, FINCH2014299, GoldenLadder, PhysRevB.93.165128, PhysRevB.93.035124, AnyonLadderAyeni} and gapped topological phases~\cite{PhysRevLett.103.070401, AnyonSpin1Chains, PhysRevB.90.075129}, where the effective degrees of freedom, in which the Hamiltonians are described, often lie in the space of composite topological charges. For so-called non-Abelian anyons, the composite charge of two or more anyons may not be unique, which may result in interactions effectively favoring certain channels~\cite{PhysRevLett.103.110403}, similar to how the spin-singlet channel is favored for antiferromagnetic interactions in spin chains.

A fundamental question in this context pertains to the influence that the anyonic exchange statistics have on the phases that may arise and how the anyons themselves interact with each other. This question naturally leads to models that are analogues to the Hubbard model for anyons, which may serve as minimal models describing itinerant anyons with effective short-range interactions arising, e.g., from screened Coulomb repulsion.

This paper deals with this topic by studying an anyonic Hubbard model on the ladder geometry for Fibonacci anyons~\cite{IntroFibAndGC}, which are a particular example of non-Abelian anyons. The anyons are subject to hopping processes and nearest-neighbor interactions, that is, there are no effective interactions in the space of composite topological charges for which interaction strengths are difficult to motivate. Focusing on the limit of half-filling, a metallic and a Mott insulating charge-density wave (CDW) phase are found and the phase transition point is determined. In the limit of strong repulsion, an effective Hamiltonian is obtained via perturbation theory and its phase diagram is determined by extracting the central charges and studying the transfer matrix spectra~\cite{Zauner_2015}, revealing four distinct phases.

The paper is structured as follows. In Section~\ref{sec:model}, the anyonic Hubbard model is introduced and its phase transition from a metallic to a CDW phase is studied. From Section~\ref{sec:eff_model} onwards, we focus on the strong repulsion limit, for which the effective Hamiltonian is introduced in this section. The numerical results of the effective model are discussed in detail in Sec.~\ref{sec:results}, where the central charges and the transfer matrix spectra are in focus. Finally, in Section~\ref{sec:conclusion}, we conclude by summarizing our findings and giving an outlook on possible future research directions.

\section{Itinerant Ladder Model}
\label{sec:model}

\begin{figure}[t]
\centering
\includegraphics{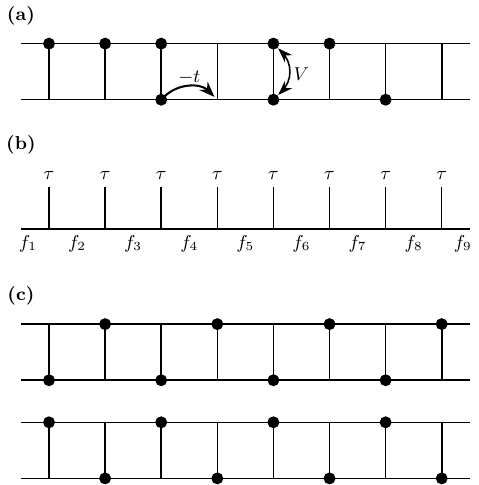}
\caption{\textbf{(a)} Ladder on which Fibonacci anyons (black dots) hop with amplitude $-t$ to the nearest neighboring (unoccupied) sites. Neighboring anyons experience repulsive interactions of strength $V$. \textbf{(b)} States in the fusion space are specified by anyonic charges $f_i\in \lbrace 1,\tau \rbrace$, where the vertices correspond to fusion processes. Each state in the full Hilbert space consists of a real space component describing the positions of the anyons on the lattice and a fusion space component (depicted here) describing how the localized anyons fuse with each other. \textbf{(c)} For $V/t\gg 1$ and half-filling, the Fibonacci anyons form a CDW in real space, of which there are two different realizations. In either case, the effective degrees of freedom lie in the fusion space.}
\label{fig:Ladder}
\end{figure}

The physical model of interest is inspired by an extended Hubbard model~\cite{hubbard} and describes itinerant Fibonacci anyons~\cite{IntroFibAndGC} on a two-leg ladder of length $L$ subject to screened Coulomb interaction. The Hamiltonian thus consists of tight-binding hopping processes and nearest-neighbor repulsion, as depicted in Fig.~\ref{fig:Ladder}\textbf{a},
\begin{align}
	\mathcal{H} &= -t\sum_{\langle i,j \rangle} \left(T_{ij}+ \text{H.c.}\right) +V\sum_{\langle i,j\rangle} n_in_j.
	\label{eq:HLadder}
\end{align}
Here, $t$ denotes the hopping amplitude, $V$ the repulsion strength and $n_i$ the occupation operator for site $i$ and our focus in the following is on the case of half-filling $\sum_i \langle n_i\rangle = L$, where $\langle \cdot \rangle$ denotes the expectation value. The sums in Eq.~\eqref{fig:Ladder} run over all nearest-neighboring pairs of lattice sites $\langle i,j \rangle$ and the operator $T_{ij}$ translates Fibonacci anyons located at site $i$ to site $j$ while incorporating the nontrivial action of the anyonic exchange processes on the wave function. {The operators $T_{ij}$ cannot be expressed in terms of simple creation and annihilation operators as for bosons and fermions due to the presence of the fusion space that is discussed below.} For details regarding the implementation of these operations in simulations, see Refs.~\cite{PhysRevB.43.2661, PhysRevB.43.10761, PhysRevB.48.13742} for the special case of Abelian anyons and Refs.~\cite{PhysRevB.107.195129, darragh_thesis, nico_master} for a general algorithm that is also applicable to non-Abelian anyons and, in particular, Fibonacci anyons. Note that due to the nontrivial exchange statistics, multiple anyons are not allowed to occupy the same site (cf. Pauli exclusion for fermions).

Fibonacci anyons are a particular type of non-Abelian anyon with the property that the composite topological charge of two Fibonacci anyons can be topologically neutral or again carry the charge of a Fibonacci anyon. This property is denoted as $\tau \times \tau = 1+\tau$, where $\tau$ refers to the topological charge of a Fibonacci anyon, $1$ to the neutral (trivial) topological charge and $\times$ to the formation of a composite charge, which is also known as fusion. A physical state then necessarily features a component specifying the details on how the anyons in the system fuse with each other, usually in diagrammatic form as depicted in Fig.~\ref{fig:Ladder}\textbf{b}. The anyonic charges appearing in these diagrams need to respect the fusion properties mentioned above in each of the vertices, that is, if $f_1=\tau$ in Fig.~\ref{fig:Ladder}\textbf{b}, then both $f_2=1$ and $f_2=\tau$ is consistent, whereas if $f_1=1$, only $f_2=\tau$ is possible due to the property $1\times \tau = \tau\times 1= \tau$. These anyonic charges determine how exactly exchange processes act on the wave function. We refer the reader to App.~\ref{App:Anyons} or, e.g., Refs.~\cite{topologicalquantum, 1506.05805, bonderson_2007, 2102.05677, InsideOutsideBases, PhysRevB.107.195129, darragh_thesis, 0707.4206, kong2022invitationtopologicalorderscategory} for an introduction to the formalism used for describing anyons. Note that fusion can in principle also occur in processes that involve anyons located at neighboring sites combining and fusing into their common channels, such that the total number of anyons is not conserved. Such processes are not considered here, which may be motivated by associating a comparably large mass $m$, with $m \gg t$ and $m \gg V$, to the anyons, effectively leading to the total anyon number being conserved. For anyons that also carry electric charge, such as the ones originating from fractional quantum Hall states~\cite{PhysRevB.87.085106, PhysRevB.95.115136}, large masses can be achieved by utilizing appropriate Coulombic charging energies~\cite{PhysRevB.87.085106}.

Note that apart from the Hamiltonian in Eq.~\eqref{eq:HLadder} that is studied here,  other anyonic Hubbard models that are described by similar Hamiltonians but feature some crucial differences have been considered in the recent literature. While some papers focus on Abelian anyons and are motivated from ultracold atom experiments, where multiple anyons are allowed to occupy the same site~\cite{PhysRevLett.126.163201, PhysRevB.108.155134, bonkhoff2024anyonicphasetransitions1d}, other works focus on (mobile or immobile) non-Abelian anyons on chain~\cite{IntroFibAndGC, GoldenChain, GoldenChainAndMG, PhysRevB.90.075129, anyonic_defects, PhysRevLett.103.070401, AnyonSpin1Chains, YangLeeAnyons, PhysRevLett.108.207201, PhysRevB.87.085106, PhysRevB.90.081111, FINCH2014299} or ladder geometries~\cite{GoldenLadder, YangLeeAnyons, PhysRevB.93.165128, PhysRevB.93.035124, AnyonLadderAyeni} and directly propose effective interactions in the fusion space~\cite{PhysRevLett.103.110403} (i.e., interactions acting on the diagrams Fig~\ref{fig:Ladder}\textbf{b}) to work with.

In contrast to the latter, while we also focus on a particular example of itinerant non-Abelian anyons, we do not propose any effective interactions in the fusion space. Instead, we focus on real space repulsion, which can be motivated as modeling the screened Coulomb interaction of electrically charged Fibonacci anyons that are expected to be supported in fractional quantum Hall systems~\cite{PhysRevB.87.085106, PhysRevB.95.115136}. Further, we use the model in Eq.~\eqref{eq:HLadder} as a starting point for deriving an effective model in Sec.~\ref{sec:eff_model} using perturbation theory. All considered interactions in the fusion space, including the interaction strengths, and the resulting phases are therefore physically motivated from hopping processes and Coulomb interaction.

\begin{figure}[t]
    \centering
    \includegraphics[]{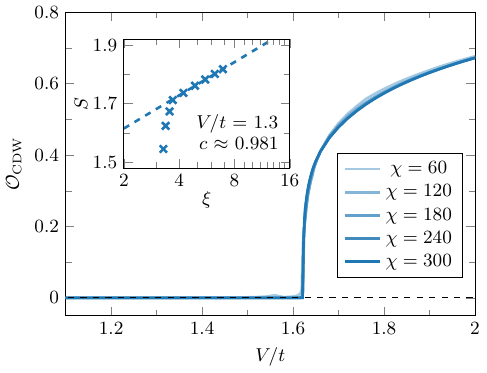}
    \caption{CDW order parameter $\mathcal{O}_{\text{CDW}}$ for the ladder Hamiltonian in Eq.~\eqref{eq:HLadder} obtained from iMPS simulations with bond dimensions of $\chi=60, 120, 180, 240, 300$. The phase transition between the metallic phase and the CDW occurs at approximately $V/t\approx 1.62$. Inset: The scaling of the entanglement entropy with correlation length for $V/t = 1.3$ yields a central charge of $c=0.981\pm 0.009$.}
   \label{fig:Ladder_transition}
\end{figure}

Intuitively, when the hopping term in the Hamiltonian in Eq.~\eqref{eq:HLadder} dominates, i.e., for $V/t \ll 1$, we expect the system to feature a metallic phase. Similarly, for $V/t \gg 1$, the system is expected to enter a Mott insulating phase, in which the ground state exhibits a charge-density wave (CDW) in real space, as shown in Fig.~\ref{fig:Ladder}\textbf{c}, where one of the two depicted CDWs is spontaneously realized. The value of $V/t$ at which the phase transition between the metallic phase and the CDW occurs can be determined using the order parameter
\begin{align}
	\mathcal{O}_{\text{CDW}}&= |\langle n_{(\pi, \pi)^\top}\rangle|,\\
	\text{with}\quad \langle n_{\mathbf{q}}\rangle &= \frac{1}{L}\sum_{j} e^{i\mathbf{r}_j\cdot \mathbf{q}} \langle n_j\rangle,
\end{align}
where $\mathbf{r}_j$ denotes the position of site $j$. The first (second) component in $\mathbf{r}_j$ indicates which rung (leg) is associated with site $j$, corresponding to the $x$ ($y$) coordinate. Numerical results for the order parameter $\mathcal{O}_{\text{CDW}}$ obtained using infinite matrix product states (iMPS)~\cite{MPS1, MPS2, MPS3} for anyonic systems~\cite{PhysRevB.82.115126, PhysRevB.89.075112, PhysRevB.92.115135, PhysRevB.93.165128} are depicted in Fig.~\ref{fig:Ladder_transition} up to bond dimension $\chi = 300$ and show that $\mathcal{O}_{\text{CDW}}$ becomes nonzero at $V/t \approx 1.62$. The system thus features a CDW in real space for $V/t \gtrsim 1.62$ ($\mathcal{O}_{\text{CDW}} \neq 0$) and a metallic phase for $V/t \lesssim 1.62$ ($\mathcal{O}_{\text{CDW}} \rightarrow 0$).

A second signature of a metallic phase is the expected value of $c=1$ for the central charge, similar to free bosons or fermions in one dimension. Indeed, it has been confirmed in Refs.~\cite{PhysRevLett.108.207201, PhysRevB.87.085106, AnyonLadderAyeni, PhysRevB.93.165128} that the \emph{kinetic} contribution to the central charge originating from itinerant (Fibonacci) anyons is $c=1$, although these works studied Hamiltonians with additional interactions in the fusion space that further modify the central charge. For a system described by a conformal field theory (CFT)~\cite{francesco2012conformal} with central charge $c$, the value of $c$ can be estimated by using iMPS to compute the entanglement entropies $S$ and correlation lengths $\xi$ across different bond dimensions and fitting them to the expected behavior~\cite{HOLZHEY1994443, Pasquale_Calabrese_2004, PhysRevB.78.024410}
\begin{align}
	S = \frac{c}{6}\ln(\xi) + \text{const}.
	\label{eq:S_scaling}
\end{align}
The corresponding numerical results for $V/t=1.3$ up to bond dimension $\chi=300$ are shown in the inset in Fig.~\ref{fig:Ladder_transition}, with the extracted central charge $c=0.981\pm 0.009$ being close to the expected value $c=1$. The results are thus consistent with the system being in a metallic phase with central charge $c=1$ for $V/t \lesssim 1.62$.

\section{Effective One-Dimensional Model: Strong Interaction Limit}
\label{sec:eff_model}

Having confirmed that the Hamiltonian in Eq.~\eqref{eq:HLadder} supports a CDW for $V/t \gtrsim 1.62$ in the previous section, we now focus on the effective Hamiltonian in the strong interaction limit $t/V \ll 1$ in order to further study the phases that may be realized by the ladder Hamiltonian. For this purpose, a perturbation theory treatment up to sixth order is employed, resulting in the effective Hamiltonian in fusion space
\begin{align}
	\widetilde{\mathcal{H}}_{\text{eff}} =& J_{\text{gc}}\mathcal{H}_{\text{gc}} +J_{\text{ring}}\mathcal{H}_{\text{ring}} + \text{const},\label{eq:Heff}\\
	J_{\text{gc}} =& \frac{1}{2}\frac{t^4}{V^3}-\left(\frac{19}{4}+ \frac{103}{27}\phi^{-1}\right)\frac{t^6}{V^5},\\
	\mathcal{H}_{\text{gc}} =& \sum_i \mathcal{P}^{i,i+1}_1,\qquad
	J_{\text{ring}}=- \frac{103}{108}\phi^2\frac{t^6}{V^5},\\
	\begin{split}
	\mathcal{H}_{\text{ring}} =& \sum_{i} \big( e^{-6\pi i/5}\mathcal{P}^{2i,2i+1}_1\mathcal{P}^{2i+1,2i+2}_1 \\ &\quad+ e^{6\pi i/5}\mathcal{P}^{2i+1,2i+2}_1\mathcal{P}^{2i+2,2i+3}_1 + \text{H.c.}\big),
	\end{split}
\end{align}
where $\phi = (1+\sqrt{5})/2$ denotes the golden ratio. The projector $\mathcal{P}^{i,i+1}_1$ acts on the Fibonacci anyons located in the $i$-th and $(i+1)$-th rung of the ladder and projects them onto their trivial fusion channel. The derivation of $\widetilde{\mathcal{H}}_{\text{eff}}$ is provided in App.~\ref{App:PT}.

The contribution $\mathcal{H}_{\text{gc}}$ to $\widetilde{\mathcal{H}}_{\text{eff}}$ is known in the literature as ``golden chain''~\cite{GoldenChain, IntroFibAndGC} and was found to realize the tricritical Ising CFT~\cite{FRIEDAN198537, francesco2012conformal} with central charge $c=7/10$ for negative prefactors and the three-state Potts CFT~\cite{DOTSENKO198454, francesco2012conformal} with $c=4/5$ for positive prefactors. The second contribution, $\mathcal{H}_{\text{ring}}$, originates from exchanging three Fibonacci anyons in both a clockwise and counter-clockwise manner (cf. Fig.~\ref{fig:PerturbationTheory}\textbf{b} in App.~\ref{App:PT}) and features a two-site unit cell. 
While this term has not been examined in the literature yet, its uniform counterpart (without the staggered phases) has been studied in Ref.~\cite{GoldenChainAndMG}. The connection between our effective Hamiltonian and the model considered in Ref.~\cite{GoldenChainAndMG} is addressed in more detail in App.~\ref{app:connection}.
%In this section of the appendix, we further explore the connection between the phases of the effective Hamiltonian discussed in Sec.~\ref{sec:results} below and the phases found in Ref.~\cite{GoldenChainAndMG}.
Note that $J_{\text{gc}}$ changes sign at $V/t\approx 3.77$, implying that both of the phases described by the golden chain Hamiltonian $\mathcal{H}_{\text{gc}}$ and additional, potentially unknown phases originating from the interactions in $\mathcal{H}_{\text{ring}}$ may be realized by the effective Hamiltonian $\widetilde{\mathcal{H}}_{\text{eff}}$. This does however not mean that all the phases to be discussed below are actually exhibited by the original ladder Hamiltonian given in Eq.~\eqref{eq:HLadder} since some of them may not lie within the actual strong interaction limit.

For convenience, we rescale the effective Hamiltonian by $|J_{\text{ring}}|$ to obtain
\begin{align}
		\mathcal{H}_{\text{eff}} =& J_{\text{eff}}\mathcal{H}_{\text{gc}} -\mathcal{H}_{\text{ring}} + \text{const},\label{eq:Heff2}
\end{align}
with $J_{\text{eff}} = J_{\text{gc}} / |J_{\text{ring}}|$, such that $J_{\text{eff}}$ can be used as tuning parameter between the different phases realized by the effective Hamiltonian.

In principle, there can always be terms appearing beyond the sixth order in perturbation theory that drive phase transitions that cannot be described by $\mathcal{H}_{\text{eff}}$. From symmetry considerations, it follows that the sixth-order correction is a natural point to stop. While $\mathcal{H}_{\text{gc}}$ preserves translation symmetry, time-reversal symmetry and spatial reflection symmetry, $\mathcal{H}_{\text{ring}}$ only preserves translations by two sites and the combination of time-reversal and reflection symmetry. The latter are precisely the symmetries one would expect from the original Hamiltonian in Eq.~\eqref{eq:HLadder} in a charge-density wave phase and thus marks the highest order we consider.\footnote{There is an additional non-invertible topological symmetry for periodic boundary conditions corresponding to encircling the flux through the ladder with a Fibonacci anyon~\cite{GoldenChain, AnyonSpin1Chains}. Such symmetries are present in every (quasi) one-dimensional system with anyonic degrees of freedom and cannot be broken.}

\section{Phase Diagram of the Effective Model}
\label{sec:results}

\begin{figure}
    \centering
    \includegraphics[width=.495\textwidth]{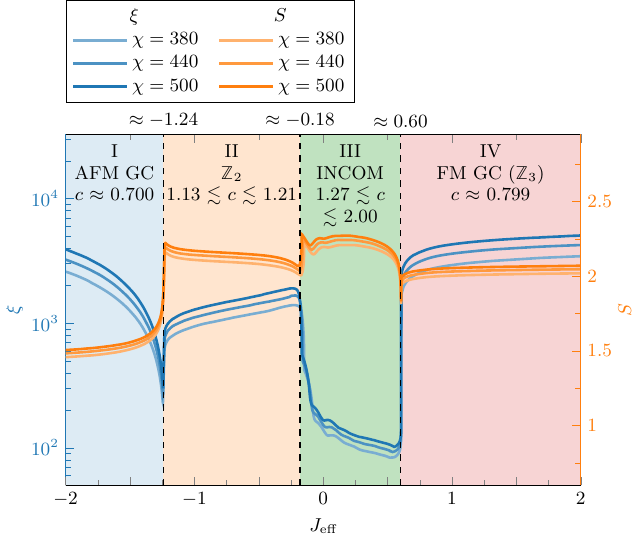}
    \caption{Phase diagram of the effective Hamiltonian in Eq.~\eqref{eq:Heff2}. The singularities in the entanglement entropy and correlation length for bond dimensions $\chi=380,440,500$ can be used to determine the phase transitions; the properties of the phases themselves are described in terms of the numerically extracted central charges and the transfer matrix spectra. Phase I: ``antiferromagnetic'' golden chain phase with $c\approx 0.700$. Phase II: the central charge can only be estimated to be within the range $1.13 \lesssim c \lesssim 1.21$; the transfer matrix indicates $\mathbb{Z}_2$ structure. Phase III: the central charge is found to be within $1.27 \lesssim c \lesssim 2.00$; the transfer matrix spectra indicate incommensurability. Phase IV: ``ferromagnetic'' golden chain phase with $c\approx 0.799$.}
   \label{fig:phase_diagram}
\end{figure}

In order to determine the phase diagram of the effective Hamiltonian in Eq.~\eqref{eq:Heff2}, we use the method of iMPS. Due to the presence of a two-site unit cell in $\mathcal{H}_{\text{eff}}$, the iMPS was also chosen to have a two-site unit cell in all simulation runs. Using these simulations, we extract the central charge $c$ using the expected entanglement scaling in Eq.~\eqref{eq:S_scaling} (Sec.~\ref{sec:centralcharge}) and analyze the spectra of the transfer matrices (Sec.~\ref{sec:TMS}). The numerical results are summarized in the phase diagram in Fig.~\ref{fig:phase_diagram}, which additionally shows the entanglement entropies and correlation lengths for bond dimensions up to $\chi=500$. The singularities in these properties are used in order to determine the phase transition points. More detailed discussions on the central charges and transfer matrix spectra can be found in the respective sections below.

We find four distinct phases: Phases I ($J_{\text{eff}} \lesssim -1.24$) and IV ($J_{\text{eff}} \gtrsim 0.60$) correspond to the tricritical Ising CFT with $c=7/10$ and the three-state Potts CFT with $c=4/5$, respectively, and are also known as ``antiferromagnetic'' golden chain (``AFM GC'') and ``ferromagnetic'' golden chain (``FM GC'') since they are realized by $\mp\mathcal{H}_{\text{gc}}$~\cite{GoldenChain, IntroFibAndGC}. For phases II ($-1.24 \lesssim J_{\text{eff}} \lesssim -0.18$) and III ($-0.18 \lesssim J_{\text{eff}} \lesssim 0.60$), the central charge can only be estimated to be within certain ranges; the transfer matrix spectra indicate that phase II features a $\mathbb{Z}_2$ structure and phase III incommensurate correlations.

We show in App.~\ref{app:connection} that all four observed phases are adiabatically connected to phases found in Ref.~\cite{GoldenChainAndMG}. This observation indicates the stability of these phases across the related models and implies that our findings complement those in Ref.~\cite{GoldenChainAndMG}, which mainly focuses on two other gapped phases as well as their critical points, neither of which has counterparts in our model. The connection in particular suggests that the observed incommensurability of correlations in phase III is not an effect arising from the staggered interaction in the effective Hamiltonian Eq.~\eqref{eq:Heff2}, but rather seems to directly originate from the three-anyon interactions. This suggests that incommensurate correlations in (non-Abelian) anyonic systems may generally be realized when considering subleading interactions involving three or more anyons.

Note that an oscillation in the entanglement entropy is observed when approaching the phase transition between phases II and III from larger $J_{\text{eff}}$. This oscillation seems to be a pure finite bond dimension effect, meaning that its position seems to converge to the phase transition point. It is nevertheless possible that in the limit of infinite bond dimensions, the position of this oscillation does not agree with the phase transition point, which would then hint at the existence of a fifth phase that we are unable to resolve numerically.

\subsection{Central Charge}
\label{sec:centralcharge}

\begin{figure}
    \centering
    \includegraphics[width=.46\textwidth]{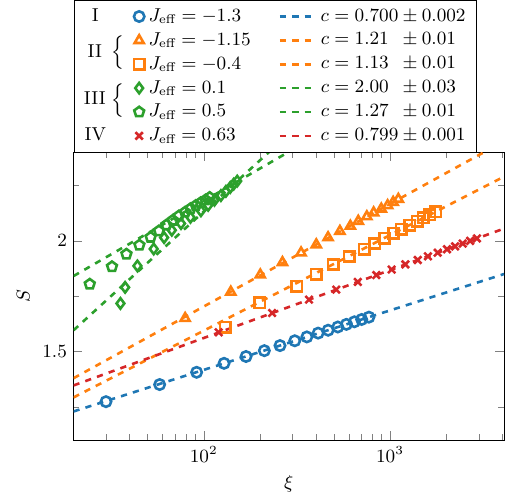}
    \caption{Extraction of central charges of the effective Hamiltonian in Eq.~\eqref{eq:Heff2} from the scaling behavior of the entanglement entropy with correlation length (see Eq.~\eqref{eq:S_scaling}) using results obtained from iMPS simulations with maximum bond dimensions ranging from $\chi=50$ to $\chi=500$, where only the last few data points were considered for the fits.}
   \label{fig:centralcharge}
\end{figure}

The central charges indicated in the phase diagram in Fig.~\ref{fig:phase_diagram} are obtained by fitting the numerical values for the entanglement entropy and correlation length to Eq.~\eqref{eq:S_scaling}. Examples of such fits are depicted in Fig.~\ref{fig:centralcharge}, where the bond dimension was varied from $\chi=50$ up to $\chi = 500$. For phase I, the data for $J_{\text{eff}}=-1.3$ is shown, for which a central charge of $c=0.700\pm 0.002$ can be extracted. This is in perfect agreement with the expectation of phase I corresponding to the tricritical Ising CFT with $c=7/10$~\cite{FRIEDAN198537, francesco2012conformal}.

For phase II, while the central charge itself can be extracted for different values of $J_{\text{eff}}$, it is found that the extracted values slowly change when tuning $J_{\text{eff}}$ within the phase. More concretely, for $J_{\text{eff}}=-1.15$, a value of $c=1.21 \pm 0.01$ is extracted, whereas for $J_{\text{eff}}=-0.4$, one finds $c=1.13 \pm 0.01$. This behavior does not allow for the extraction of an unambiguous value for the central charge, such that only the interval $1.13 \lesssim c \lesssim 1.21$ can be roughly estimated. The main reason for this is most probably that the simulations need larger bond dimensions to converge to a single value for the central charge across all $J_{\text{eff}}$ in this phase. One would thus expect the interval to shrink and maybe slightly shift with higher bond dimensions and eventually converge to a single value. Note that there is also the possibility of this phase not having a well-defined central charge even in the limit of infinite bond dimensions. Such a case cannot be ruled out based on our numerics, meaning that our results for the central charge are in fact inconclusive. The same holds for phase III discussed below.

For phase III, the convergence of the numerics to the fixed point value of the central charge is found to be similarly slow as in phase II, yielding the interval $1.27 \lesssim c \lesssim 2.00$ as a rough estimate for the central charge. For $J_{\text{eff}}=0.1$, for example, a value of $c=2.00 \pm 0.03$ is found, whereas for $J_{\text{eff}}=0.5$, $c=1.27 \pm 0.01$.

Finally, for phase IV, Fig.~\ref{fig:centralcharge} shows the numerical results and the corresponding fit for $J_{\text{eff}}=0.63$, with the extracted value for the central charge being $c=0.799\pm 0.001$. This value is in excellent agreement with the expected central charge $c=4/5$ of the three-state Potts CFT that is expected to describe this phase~\cite{DOTSENKO198454, francesco2012conformal}.

\subsection{Transfer Matrix Spectra}
\label{sec:TMS}

\begin{figure}
    \centering
    \includegraphics[]{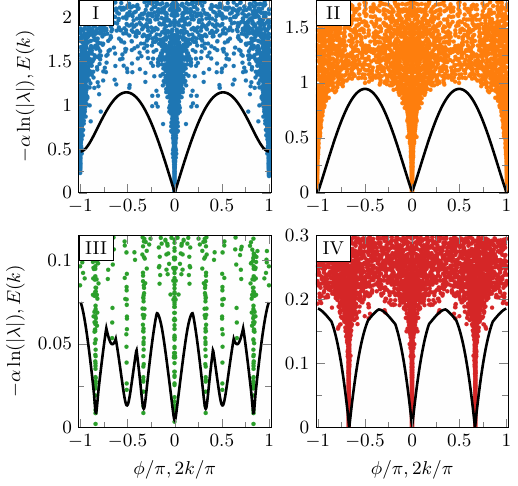}
    \caption{Transfer matrix spectra of the effective Hamiltonian $\mathcal{H}_{\text{eff}}$ obtained from iMPS simualtions using two-site unit cells and bond dimension $\chi=100$ plotted with respect to the rescaled logarithm of the absolute values, $-\alpha \ln(|\lambda|)$, and the complex phases $\phi=\arg(\lambda)$ for $J_{\text{eff}}=-1.6$ (phase I, upper left, $\alpha=0.37$), $J_{\text{eff}}=-0.8$ (phase II, upper right, $\alpha=0.24$), $J_{\text{eff}}=0.5$ (phase III, lower left, $\alpha=0.13$) and $J_{\text{eff}}=0.8$ (phase IV, lower right, $\alpha=0.068$). The solid lines correspond to the lowest-lying excitation $E(k)$ for each $J_{\text{eff}}$.}
   \label{fig:TMS_phases}
\end{figure}

It has been shown in Ref.~\cite{Zauner_2015} that the spectra of iMPS transfer matrices are related to the dispersions of the low-lying excitations of the system. In particular, it has been found that the complex phases of the most dominant eigenvalues (i.e., the eigenvalues with the largest magnitudes) essentially correspond to the momenta for which the dispersions of the low-lying excitations are minimal. For gapless phases, such as the ones that are considered here, these complex phases can thus be used to find the momenta for which the system becomes gapless~\cite{Zauner_2015}. In Fig.~\ref{fig:TMS_phases}, the transfer matrix spectra are shown across all four distinct phases for bond dimension $\chi =  100$ in terms of their magnitudes $-\ln(|\lambda|)$ and complex phases $\phi=\arg(\lambda)$. The magnitudes $-\ln(|\lambda|)$ have been rescaled by an additional factor $\alpha$ such that the values are comparable to the dispersions of the lowest-lying excitations obtained from a quasiparticle ansatz for MPS~\cite{PhysRevB.55.2164, PhysRevB.85.100408, PhysRevB.88.075133, 10.21468/SciPostPhysLectNotes.7}, which are shown as solid lines. Note that due to the two-site unit cell of the effective Hamiltonian and the iMPS, the momenta $k$ in Fig.~\ref{fig:TMS_phases} are rescaled by $\pi/2$, while the phases $\phi$ of the transfer matrix spectra are rescaled by $\pi$.

Across all plots, it can be observed that the rescaled momenta associated with the low-lying dispersions' minima and, in particular, the gapless points indeed agree with the rescaled complex phases of the transfer matrix spectra. For phase I (upper left plot in Fig.~\ref{fig:TMS_phases}), there is only a single gapless point at $k=0$. This agrees with the expectation of this phase realizing the tricritical Ising CFT. The operator content of this CFT only contains fields that are symmetric under translations by one or two sites (corresponding to momenta $k=0$ and $k=\pi$)~\cite{FRIEDAN198537, GoldenChain}, which implies that both momenta correspond to $k=0$ when considering a two-site unit cell, as done here.

For phase II (upper right plot in Fig.~\ref{fig:TMS_phases}), two gapless points at $k=0$ and $k=\pi/2$ are found. Due to the presence of the two-site unit cell, it is possible for this phase to be gapless at the momenta $k=n\pi/2$, $n\in \mathbb{Z}_4$, if it can be adiabatically connected to a translational invariant Hamiltonian realizing the same phase. This is indeed the case, as shown in App.~\ref{app:connection}.

Phase III (lower left plot in Fig.~\ref{fig:TMS_phases}) is observed to feature incommensurability in its correlations, that is, the momenta associated with the minima of the lowest-lying excitation dispersion seem to change continuously. In order to make this behavior more apparent, the transfer matrix spectra corresponding to additional values of $J_{\text{eff}}$ are plotted in Figs.~\ref{fig:TMS1} and \ref{fig:TMS2} in App.~\ref{App:TMS}.

For Phase IV (lower right plot in Fig.~\ref{fig:TMS_phases}), the three gapless points at $k=0,\pm \pi/3$ are consistent with the expectation of this phase being described by the three-state Potts CFT. The presence of the $\mathbb{Z}_3$ symmetry~\cite{DOTSENKO198454} implies that gapless points may be found at momenta $k=0,\pm 2\pi/3$ for a single-site unit cell, corresponding to $k=0,\mp \pi/3$ for a two-site unit cell.\\ 

We emphasize that the transfer matrix spectra indicate what kinds of \emph{correlations} can be found within each of the four phases. This is not a statement about whether or not the phases are ordered. For phase IV for example, we observe the $\mathbb{Z}_3$ structure in the transfer matrix eigenvalues, which is reflected in the behavior of algebraically decaying correlations between the corresponding $\mathbb{Z}_3$ primary fields of the three-state Potts CFT~\cite{DOTSENKO198454}. Similarly, phases II and III must feature correlations with $\mathbb{Z}_2$ and incommensurate behavior, respectively. An example for an operator for which such a behavior can be observed in our case corresponds to the local density of trivial charges in fusion space~\cite{GoldenChainAndMG}, or, equivalently, to the local projector $\mathcal{P}^{i,i+1}_1$.

\section{Conclusion}
\label{sec:conclusion}

We have studied Fibonacci anyons localized on a ladder subject to a tight-binding model with nearest-neighbor interactions and have analyzed the supported phases at half-filling. For low repulsion strengths, a metallic phase with central charge $c=1$ is found, which arises from the hopping processes in the Hamiltonian. This is consistent with results from other works~\cite{PhysRevLett.108.207201, PhysRevB.87.085106, AnyonLadderAyeni, PhysRevB.93.165128} in which the total central charge is found to consist of this kinetic contribution and fusion space contributions that are trivial in our case. For larger repulsion strengths, the system enters a charge-density wave (CDW) phase, which is confirmed with the appropriate order parameter.

Employing a perturbation theory treatment in the strong repulsion limit, it is found that the resulting effective Hamiltonian supports four distinct phases. Two of these phases are described by the tricritical Ising CFT and the three-state Potts CFT, which are known as the phases realized by the golden chain Hamiltonian~\cite{GoldenChain, IntroFibAndGC}. The presence of these phases is confirmed by extracting consistent central charges and transfer matrix spectra of infinite matrix product states. Using these properties to characterize the remaining two phases, one phase is found to exhibit incommensurate correlations while the other one supports $\mathbb{Z}_2$ behavior. In both phases, the convergence with bond dimension is too slow in order to unambiguously extract the central charge, such that only intervals in which the true values may lie can be estimated. All four phases of the effective Hamiltonian are shown to be adiabatically connected to phases of a known anyon model~\cite{GoldenChainAndMG}.

Considering the above results, there are many interesting questions that may be worth exploring in future research. As it was found that the three-anyon interactions lead to incommensurate correlations for Fibonacci anyons, it would be interesting to consider general three-anyon interactions (not motivated from our effective Hamiltonian) for other non-Abelian anyon species and explore how generic such incommensurate correlations are for these models.

Another prospect is the generalization of our model to other anyon types and the search for phases that similarly cannot be described by anyonic chain Hamiltonians that only feature nearest-neighbor interactions~\cite{GoldenChain, IntroFibAndGC, PhysRevLett.103.070401, AnyonSpin1Chains, YangLeeAnyons, PhysRevB.90.081111, FINCH2014299}. For such studies, the perturbation theory calculation presented in App.~\ref{App:PT} can be straight forwardly adapted to other anyon types.

Even when staying with Fibonacci anyons, going beyond the case of half-filling seems to be a promising direction. In particular, as there have already been studies on anyonic analogues to the $t-J$ model~\cite{PhysRevLett.108.207201, PhysRevB.87.085106, PhysRevB.93.165128, AnyonLadderAyeni} focusing on nearest-neighbor interactions, employing perturbation theory to derive an effective Hamiltonian that contains contributions beyond these analogues may reveal new phases and insights, similar to how the effective model studied in this work supports phases beyond the golden chain.

\section*{Acknowledgements}

N.K. and F.P. thank Fabian Essler, Paul Fendley, and Steve Simon for insightful discussions and pointing out the connection between our effective Hamiltonian and the model discussed in Ref.~\cite{GoldenChainAndMG}.
The numerical simulations have been carried out using the MPSKit library~\cite{MPSKit}.
N.K. and F.P. acknowledge support from the European Research Council (ERC) under the European Union’s Horizon 2020 research and innovation program under Grant Agreement No. 771537, the Deutsche Forschungsgemeinschaft (DFG, German Research Foundation) under Germany’s Excellence Strategy EXC-2111-390814868, TRR 360 - 492547816, FOR 5522 (project-id 499180199), and the Munich Quantum Valley, which is supported by the Bavarian state government with funds from the Hightech Agenda Bayern Plus.
R.M. acknowledges support from the Deutsche Forschungsgemeinschaft (DFG, German Research Foundation) via the cluster of excellence ct.qmat (EXC-2147, project-id 390858490) and SFB 1143 (project-id 247310070).
A.G-S. acknowledges support from the Royal Commission for the Exhibition of 1851, and support from the UK Research and Innovation (UKRI) under the UK government’s Horizon Europe funding guarantee [grant number EP/Y036069/1].

\section*{Data Availability}

Access to the simulation codes and numerical data shown in the figures may be granted upon reasonable request~\cite{zenodo}.

\appendix

\section{Basics of Anyon Models}
\label{App:Anyons}

In this section, we give a short introduction to the formalism that is used for describing anyons in general and discuss the concrete case of Fibonacci anyons. Only the most elementary aspects that are needed in order to understand the effective model and its derivation in App.~\ref{App:PT} are covered. For more elaborate introductions to this topic, see, e.g., Refs.~\cite{topologicalquantum, 1506.05805, bonderson_2007, 2102.05677, InsideOutsideBases, PhysRevB.107.195129, darragh_thesis}, on which this section is based.

From an abstract point of view, anyon models are specified using four different properties. The first one is the set $\mathcal{C}$ of topological charges, i.e., the set of anyons. This set always features a trivial, topologically neutral charge $1$ and for each charge $a\in \mathcal{C}$, there is also an anticharge $\overline{a}\in \mathcal{C}$. In the case of Fibonacci anyons, we have $\mathcal{C}=\lbrace 1 , \tau\rbrace$, where $\tau$ denotes the Fibonacci anyons with $\overline{\tau} = \tau$. Secondly, every anyon theory comes with a set of $N$-symbols specifying fusion, that is, the composite charges of systems with more than a single anyon. Written as
\begin{align}
	a \times b = \sum_c N_{ab}^c  c,
\end{align}
a value $N_{ab}^c\neq 0$ signifies that anyons $a$ and $b$ may be in a state such that their composite topological charge is $c$; $N$-symbols with $N_{ab}^c> 1$ indicate that there are multiple inequivalent ways of forming composite charge $c$ from $a$ and $b$. Fusion is both commutative and associative and we have $N_{ab}^c \in \mathbb{N}_0$. Due to $1$ being the trivial charge, fusion with it must also be trivial, which implies that $N_{ab}^1 = \delta_{b\overline{a}}$ and $N_{a1}^b = N_{1a}^b = \delta_{ab}$. An anyon $a$ is called non-Abelian if there is another anyon $b$ such that their fusion outcome is not unique, i.e., if $\sum_cN_{ab}^c >1$. One example for non-Abelian anyons are Fibonacci anyons, which obey $\tau \times\tau = 1+\tau$, such that $N_{\tau\tau}^1=N_{\tau\tau}^\tau=1$; all other $N$-symbols are trivial.

A physical state in which anyons $a$ and $b$ fuse to $c$ can be represented diagrammatically as
\begin{align}
	\begin{split}
	\vcenter{\includegraphics{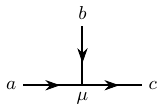}}\hspace{-155pt},
	\end{split}
\end{align}
where the directed lines are labeled by their topological charges and the vertex by the fusion multiplicity $\mu \in \lbrace 1,\ldots,N_{ab}^c \rbrace$. States associated with more than two anyons are obtained by combining multiple trivalent diagrams of the above form (cf. Fig.~\ref{fig:Ladder}\textbf{b}). Reversing the direction of an arrow is equivalent to replacing the associated topological charge by its anticharge. For anyon theories in which each charge is its own anticharge, the arrows thus do not carry any information and may be omitted. Further, for anyon models featuring fusion multiplicities with $N_{ab}^c\in \lbrace 0, 1\rbrace$ for all $a,b,c \in \mathcal{C}$, there is only a single value $\mu$ can take, such that this label may also be omitted. Since Fibonacci anyons fulfill both conditions, we can write the diagrams as done in Fig.~\ref{fig:Ladder}\textbf{b} for convenience from now on and drop the arrows and multiplicity indices.

When considering states with, e.g., three anyons $a$, $b$ and $c$, one can choose different bases. In one basis, $a$ and $b$ may fuse to $e$ which then fuses with $c$ to $d$, while in another basis, $b$ and $c$ may fuse to $f$ which then fuses with $a$ to $d$. These two bases can be expressed diagrammatically as
\begin{align}
	\begin{split}
	\vcenter{\includegraphics{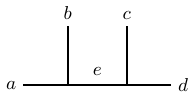}}\hspace{-134pt} \text{ and } \vcenter{\includegraphics{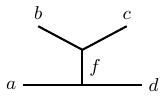}}\hspace{-155pt},
	\end{split}
\end{align}
respectively, and the corresponding basis change can be expressed in terms of $F$-symbols, which are the third of the four properties specifying an anyon theory and are defined as
\begin{align}
	\begin{split}
	\vcenter{\includegraphics{images/ThreeAnyonBasis1.pdf}}\hspace{-145pt} =\sum_f [F^{abc}_d]_{ef} \hspace{-15pt} \vcenter{\includegraphics{images/ThreeAnyonBasis2.pdf}}\hspace{-155pt}.
	\end{split}
\end{align}
Since they correspond to a change of basis, the $F$-symbols are unitary with $[(F^{abc}_d)^{-1}]_{fe}=[F^{abc}_d]_{ef}^*$. If one of the three topological charges $a$, $b$ or $c$ is the trivial charge $1$ and all charges are consistent with the fusion rules, the corresponding $F$-symbol is trivial, $[F^{abc}_d]=\mathbb{1}$. For the particular case of Fibonacci anyons, the remaining nontrivial $F$-symbols are $[F^{\tau\tau\tau}_1]_{\tau\tau}=1$ and
\begin{align}
	F^{\tau\tau\tau}_\tau = \begin{pmatrix}
	\phi^{-1} & \phi^{-1/2}\\
	\phi^{-1/2} & -\phi^{-1}
	\end{pmatrix},
\end{align}
where $\phi=(1+\sqrt{5})/2$ is the golden ratio and the first and second row / column corresponds to topological charges $1$ and $\tau$, respectively.

The final piece of information that is needed in order to fully specify anyon models are the $R$-symbols, which encode (anticlockwise) exchange phases via
\begin{align}
	\begin{split}
	\vcenter{\includegraphics{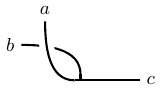}}\hspace{-155pt} =R^{ab}_c \hspace{-5pt} \vcenter{\includegraphics{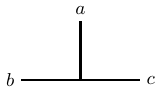}}\hspace{-160pt}.
	\end{split}
	\label{eq:Rsym}
\end{align}
Note that the exchange phase in general depends on the composite charge of the anyons being exchanged. The phases associated with the corresponding clockwise exchanges are given by $(R^{ba}_c)^*$ and exchanges with the trivial charge $1$ have no effect, $R^{a1}_a=R^{1a}_a=1$. The nontrivial $R$-symbols for Fibonacci anyons are $R^{\tau\tau}_1=e^{-4\pi i/5}$ and $R^{\tau\tau}_\tau=e^{3\pi i/5}$. When exchanging two anyons in a basis in which they do not directly fuse with each other, $F$-symbols can be used to bring the diagram to a form such that the definition of the $R$-symbols in Eq.~\eqref{eq:Rsym} can be applied. For example, we have
\begin{align}
	\begin{split}
	\vcenter{\includegraphics{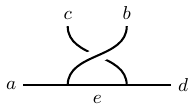}}\hspace{-145pt} =&\sum_{f,g} [F^{abc}_d]_{eg} R^{bc}_g [F^{acb}_d]_{fg}^*\hspace{-5pt}\\
	\times&\hspace{-10pt}\vcenter{\includegraphics{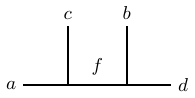}}\hspace{-142pt}.
	\end{split}
	\label{eq:braid}
\end{align}
For a chain of identical anyons, i.e., $b=c$, this can be reinterpreted as a sum over projectors of two anyons of charge $b$ onto their different fusion channels multiplied by the associated $R$-symbols, such that we may write Eq.~\eqref{eq:braid} for $b=c$ as
\begin{align}
	\begin{split}
	\vcenter{\includegraphics{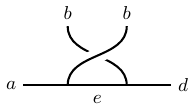}}\hspace{-145pt} =&\sum_{f,g}  [\mathcal{P}^{bb}_g]_{ef} R^{bb}_g\hspace{-5pt}\\
	\times&\hspace{-10pt}\vcenter{\includegraphics{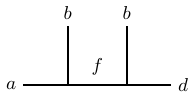}}\hspace{-142pt}
	\end{split}
	\label{eq:braid_}
\end{align}
with the matrix elements of the projectors being
\begin{align}
	[\mathcal{P}^{bb}_g]_{ef} = [F^{abb}_d]_{eg} [F^{abb}_d]_{fg}^*.
\end{align}
This relation is used in order to obtain the projectors in the effective Hamiltonian in App.~\ref{App:PT}. In particular, a sum of clockwise and anticlockwise exchanges corresponds to replacing $R^{bb}_g \rightarrow 2\text{Re}(R^{bb}_g)$ in Eq.~\eqref{eq:braid_} (cf. Eq.~\eqref{eq:CWplusACW}).

%In particular, a sum of clockwise and anticlockwise exchanges can be expressed as $2\sum_g\mathcal{P}^{bb}_g\text{Re}(R^{bb}_g)$ (cf. Eq.~\eqref{eq:CWplusACW}), where $\mathcal{P}^{bb}_g$ should now be interpreted as acting on the diagrams.

\section{Perturbation Theory}
\label{App:PT}

In this section, we derive the effective Hamiltonian in Eq.~\eqref{eq:Heff} for $t/V \ll 1$. Let us denote the full Hamiltonian in Eq.~\eqref{eq:HLadder} as
\begin{align}
	\mathcal{H} = H_V + H_t,
\end{align}
where $H_V$ denotes the Hamiltonian containing the repulsive interactions between the anyons and $H_t$ the kinetic Hamiltonian, which is treated as perturbation in the following. Further, let $P_0$ denote the projector onto the low-energy subspace of $H_V$ and $Q_0 = \mathbb{1}-P_0$. That is, $P_0$ projects onto the two CDW configurations depicted in Fig.~\ref{fig:Ladder}\textbf{c} and does not project out any fusion space configurations. Since the CDWs completely avoid nearest-neighbor occupations, the energy $E_0$ associated with $H_V$ of the states in this subspace is $E_0 = 0$. Following Ref.~\cite{PerturbationTheory} and adapting their notation, the effective Hamiltonian can be written using expressions of the form
\begin{align}
	\langle \lambda_1\lambda_2\ldots\lambda_n\rangle \equiv P_0 H_t S^{\lambda_1} H_t S^{\lambda_2} H_t \ldots H_t S^{\lambda_n} H_t P_0 \\
	\text{with}\ S^{\lambda} = 
	\begin{cases}
      P_0, & \text{if}\ \lambda=0 \\
      \frac{1}{(H_V-E_0)^{\lambda}}Q_0 = \frac{1}{H_V^{\lambda}}Q_0, & \text{otherwise,}
    \end{cases}
\end{align}
where $\lambda_i \in \mathbb{N}_0$.
Note that for $\lambda_i > 0$, it does not matter whether $Q_0$ acts from the right (as written above) or from the left since both $P_0$ and $Q_0$ commute with $H_V$.

Due to $P_0$ projecting onto the CDWs, $\langle \lambda_1\lambda_2 \ldots\lambda_n \rangle$ can only be nontrivial if the number of $H_t$ is even, that is, $n$ must be odd. This implies that only even orders in the perturbation theory contribute to the effective Hamiltonian. Further, contributions with $\lambda_1=0$, $\lambda_i=\lambda_{i+1}=0$ or $\lambda_n=0$ vanish since $P_0H_tP_0 = 0$. Following the computation in Ref.~\cite{PerturbationTheory} and incorporating these simplifications, the nonzero contributions to the effective Hamiltonian up to sixth order are
\begin{align}
	\mathcal{H}_{\text{eff}}^{(2)} &= -\langle 1 \rangle,\\
	\mathcal{H}_{\text{eff}}^{(4)} &= -\langle 111\rangle +\frac{1}{2}\big(\langle 102 \rangle + \text{H.c.}\big), \qquad \text{and}\\
	\begin{split}
	\mathcal{H}_{\text{eff}}^{(6)} &= -\langle 11111\rangle -\frac{1}{4}\langle 20102\rangle-\frac{3}{8}\big(\langle 10202\rangle+\text{H.c.}\big)\\ &\ +\frac{1}{2}\big(\langle 10211\rangle +\langle 10121\rangle +\langle 10112\rangle+ \langle 11102\rangle \\&\ -\langle 10103\rangle +\text{H.c.}\big).
	\end{split}
\end{align}
For the computation of the individual contributions at each order, we assume periodic boundary conditions and an even number of rungs, $L\ \text{mod}\ 2=0$.

\subsubsection*{Second Order}
The second-order term $\mathcal{H}_{\text{eff}}^{(2)}$ involves anyons hopping to nearest-neighboring sites and back to their initial positions and corresponds to a constant energy shift given by
\begin{align}
	\mathcal{H}_{\text{eff}}^{(2)} = -\langle 1 \rangle = -\frac{3}{2}\frac{t^2}{V}L.
\end{align}
The fraction arises due to each of the anyons having three hopping processes available (i.e., $3L$ hopping processes in total) while the eigenvalue with respect to $H_V^{-1}$ after a single hopping process being $(2V)^{-1}$.

\subsubsection*{Fourth Order}

\begin{figure}
\centering
\includegraphics{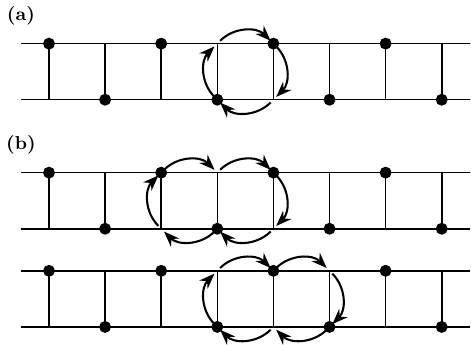}
\caption{\textbf{(a)} The exchange of two Fibonacci anyons located on neighboring rungs is a fourth-order process. The depicted clockwise exchange is denoted by $B_{CW}^{i,i+1}$, the anticlockwise counterpart appears with the same amplitude in the perturbation theory. \textbf{(b)} At sixth order, three anyons can perform a ring exchange corresponding to $B_{CW}^{i+1,i+2}B_{CW}^{i,i+1}$ (upper case) or $B_{CW}^{i,i+1}B_{CW}^{i+1,i+2}$ (lower case) for clockwise exchanges, depending on the location of the involved anyons within the CDW.}
\label{fig:PerturbationTheory}
\end{figure}

At fourth order, the term $\langle 102 \rangle$ is very similar to $\mathcal{H}_{\text{eff}}^{(2)}$ since due to the projection $S^0=P_0$,
\begin{align}
	\langle 102 \rangle = \frac{1}{2}\big(\langle 102 \rangle + \text{H.c.}\big) = \langle 1 \rangle \langle 2 \rangle = \frac{9}{8}\frac{t^4}{V^3}L^2.
	\label{eq:4thOrder1}
\end{align}
This contribution is quadratic in the system size $L$ and needs to cancel with another quadratic term in $-\langle 111\rangle$. The processes contained in $-\langle 111\rangle$ are more complicated as they involve four hopping processes until the anyons are back in their initial CDW configuration. Apart from processes in which all anyons return to their initial positions, there is now the possibility of exchanging two anyons located on neighboring rungs, as depicted in Fig.~\ref{fig:PerturbationTheory}\textbf{a}. Such contributions act nontrivially on the fusion space and we denote the operators corresponding to clockwise (anticlockwise) exchanges as $B_{CW}^{i,i+1}$ ($B_{ACW}^{i,i+1}$), where $i$ and $i+1$ specify the two associated rungs. Focusing for now on this case, the corresponding term in $-\langle 111\rangle$ reads
\begin{align}
	-\frac{1}{2}\frac{t^4}{V^3}\sum_i \left( B_{CW}^{i,i+1} + B_{ACW}^{i,i+1}\right) .
\end{align}
This can be reexpressed using the anyonic $R$-symbols and projectors onto the fusion channels of the involved anyons via
\begin{align}
	 B_{CW}^{i,i+1} + B_{ACW}^{i,i+1} = 2\sum_f \mathcal{P}^{i,i+1}_f \text{Re}\left(R^{\tau\tau}_f\right),
	 \label{eq:CWplusACW}
\end{align}
where the sum runs over all allowed fusion channels $f$ ($f=1,\tau$ for Fibonacci anyons) and $\mathcal{P}^{i,i+1}_f$ denotes the projector acting on the anyons in the $i$-th and $(i+1)$-th rung that projects them onto their $f$ fusion channel. For Fibonacci anyons, we have~\cite{bonderson_2007}
\begin{gather}
	\mathcal{P}^{i,i+1}_1 + \mathcal{P}^{i,i+1}_\tau = \mathbb{1},\\
	R^{\tau\tau}_1 = e^{-4\pi i/5}, \quad R^{\tau\tau}_\tau = e^{3\pi i/5},\\
	2\text{Re}\left(R^{\tau\tau}_1\right) = -\phi  \quad\text{and}\quad   2\text{Re}\left(R^{\tau\tau}_\tau\right) = -\phi^{-1},
\end{gather}
where $\phi=(1+\sqrt{5})/2$ denotes the golden ratio. That is,
\begin{align}
	\begin{split}
	&-\frac{1}{2}\frac{t^4}{V^3}\sum_i \left( B_{CW}^{i,i+1} + B_{ACW}^{i,i+1}\right) \\
	&= \frac{1}{2}\frac{t^4}{V^3}\sum_i \mathcal{P}^{i,i+1}_1 + \frac{1}{2}\phi^{-1}\frac{t^4}{V^3}L.
	\end{split}
	\label{eq:BCWBACW}
\end{align}

The remaining processes in $-\langle 111\rangle$ that do not involve exchanges contribute
\begin{align}
	- \frac{9}{8}\frac{t^4}{V^3}L^2 + \frac{17}{24}\frac{t^4}{V^3}L
\end{align}
to the effective Hamiltonian such that the term quadratic in the system size in Eq.~\eqref{eq:4thOrder1} indeed cancels and the fourth order of the effective Hamiltonian reads
\begin{align}
	\mathcal{H}_{\text{eff}}^{(4)} = \frac{1}{2}\frac{t^4}{V^3}\sum_i \mathcal{P}^{i,i+1}_1 + \left(\frac{17}{24}+\frac{1}{2}\phi^{-1}\right)\frac{t^4}{V^3}L.
\end{align}

\subsubsection*{Sixth Order}

For the sixth-order Hamiltonian, we start by considering contributions like $\langle 20102\rangle$ and $\langle 10211\rangle$, which feature projectors $P_0$ in the corresponding expressions. Similar to before, one can evaluate these contributions by going back to previous computations and slightly adapting them. For example, we have
\begin{align}
	\begin{split}
	&\langle \lambda_1 0 \lambda_3 0 \lambda_5 \rangle = \langle \lambda_1 \rangle \langle \lambda_3  \rangle \langle \lambda_5 \rangle = \frac{27}{32} \frac{t^6}{V^5}L^3\\
	&\text{for}\ \lambda_1 + \lambda_3 + \lambda_5 = 5.
	\end{split}
\end{align}
For contributions involving the exchange process illustrated in Fig.~\ref{fig:PerturbationTheory}\textbf{a}, we find
\begin{align}
	\langle 10211\rangle &= \langle 1\rangle \langle 211\rangle = \langle 10112\rangle = \langle 11102 \rangle,
\end{align}
\begin{align}
	\begin{split}
	\langle 10211\rangle &= \frac{3}{4}\frac{t^6}{V^5}L \bigg(\frac{9}{8}L^2 -\frac{17}{24}L \\ &\qquad\quad\quad\,\,\,\,\, +\frac{1}{2}\sum_i \left( B_{CW}^{i,i+1} + B_{ACW}^{i,i+1}\right) \bigg),
	\end{split}\\
	\begin{split}
	\langle 10121\rangle &= \frac{3}{2}\frac{t^6}{V^5}L \bigg(\frac{9}{32}L^2 +\frac{85}{288}L \\ &\qquad\quad\quad\,\,\,\,\, +\frac{1}{4}\sum_i \left( B_{CW}^{i,i+1} + B_{ACW}^{i,i+1}\right) \bigg).
	\end{split}
\end{align}
Summing all these terms together with their respective prefactors yields
\begin{align}
	\frac{81}{64}\frac{t^6}{V^5}L^3  - \frac{221}{192}\frac{t^6}{V^5}L^2  + \frac{3}{2}\frac{t^6}{V^5}L \sum_i \left( B_{CW}^{i,i+1} + B_{ACW}^{i,i+1}\right).
	\label{eq:6thOrderIntermediate}
\end{align}
Since these contributions do not scale linearly in system size, they will eventually cancel with terms in $-\langle 11111\rangle$. The processes contained in $-\langle 11111\rangle$ involve, apart from what we have already seen in the second and fourth-order contributions, processes in which three anyons perform a ``ring exchange'', as depicted in Fig.~\ref{fig:PerturbationTheory}\textbf{b}. These processes can be expressed in terms of braid operators as $B_{CW}^{i+1,i+2}B_{CW}^{i,i+1}$ and $B_{CW}^{i,i+1}B_{CW}^{i+1,i+2}$ for clockwise ring exchanges, see Fig.~\ref{fig:PerturbationTheory}\textbf{b}. Crucially, this leads to the formation of a two-site unit cell and the effective Hamiltonians corresponding to the two possible CDW configurations are related by shifting the system by a single site. Focusing on one specific CDW configuration, we can evaluate all contributions to $\langle 11111\rangle$:
\begin{align}
	\begin{split}
	\langle 11111\rangle &= \left(\frac{81}{64}L^3 - \frac{221}{192}L^2 + \frac{347}{432}L\right)\frac{t^6}{V^5}\\
	&\ + \left(\frac{3}{2}L - \frac{101}{108}\right)\frac{t^6}{V^5} \sum_i \left( B_{CW}^{i,i+1} + \text{H.c.}\right)\\
	&\ + \frac{103}{108}\frac{t^6}{V^5}\sum_i \Big( B_{CW}^{2i,2i+1}B_{CW}^{2i+1,2i+2}\\
	&\qquad\quad\quad\quad\,\,\ \ + B_{CW}^{2i+2,2i+3}B_{CW}^{2i+1,2i+2} + \text{H.c.} \Big).
	\end{split}
	\label{eq:6th_order_11111}
\end{align}
As expected, the terms scaling nonlinearly in the system size in Eq.~\eqref{eq:6thOrderIntermediate} are canceled when summed with $-\langle 11111\rangle$. The final contribution in Eq.~\eqref{eq:6th_order_11111} can be rewritten as
\begin{align}
	\begin{split}
	B_{CW}^{2i,2i+1}&B_{CW}^{2i+1,2i+2} = \phi^2e^{-6\pi i/5}\mathcal{P}^{2i,2i+1}_1\mathcal{P}^{2i+1,2i+2}_1\\
	&\,\,\,\,\,\,\,\,\,\,\, + \phi\left(\mathcal{P}^{2i,2i+1}_1 + \mathcal{P}^{2i+1,2i+2}_1\right) + e^{6\pi i/5},
	\end{split}
\end{align}
such that the full sum can be expressed as
\begin{align}
	\begin{split}
	\sum_i& \left( B_{CW}^{2i,2i+1}B_{CW}^{2i+1,2i+2}+ B_{CW}^{2i+2,2i+3}B_{CW}^{2i+1,2i+2} + \text{H.c.} \right)\\
	=&\phi^2 \sum_i \Big( e^{-6\pi i/5}\mathcal{P}^{2i,2i+1}_1\mathcal{P}^{2i+1,2i+2}_1 \\
	&+ e^{-6\pi i/5}\mathcal{P}^{2i+2,2i+3}_1\mathcal{P}^{2i+1,2i+2}_1 + \text{H.c.} \Big)\\
	&+ 4\phi\sum_i \mathcal{P}^{i,i+1}_1 -\phi L.
	\end{split}
\end{align}
Thus, together with Eq.~\eqref{eq:BCWBACW}, the sixth-order correction can be expressed as
\begin{align}
	\begin{split}
	\mathcal{H}_{\text{eff}}^{(6)} &=  -\frac{103}{108}\phi^2 \frac{t^6}{V^5} \sum_i \Big( e^{-6\pi i/5}\mathcal{P}^{2i,2i+1}_1\mathcal{P}^{2i+1,2i+2}_1 \\
	&\ + e^{6\pi i/5}\mathcal{P}^{2i+1,2i+2}_1 \mathcal{P}^{2i+2,2i+3}_1 + \text{H.c.} \Big)\\
	&\ -\left(\frac{19}{4}+\frac{103}{27}\phi^{-1}\right)\frac{t^6}{V^5}\sum_i \mathcal{P}^{i,i+1}_1\\
	&\ + \left(\frac{103}{108}\phi - \frac{347}{432} -\frac{101}{108}\phi^{-1} \right)\frac{t^6}{V^5}L,
	\end{split}
\end{align}
and the full effective Hamiltonian up to sixth order takes the form
\begin{widetext}
\begin{align}
	\begin{split}
	\mathcal{H}_{\text{eff}} &= -\frac{103}{108}\phi^2 \frac{t^6}{V^5} \sum_i \left( e^{-6\pi i/5}\mathcal{P}^{2i,2i+1}_1\mathcal{P}^{2i+1,2i+2}_1 + e^{6\pi i/5}\mathcal{P}^{2i+1,2i+2}_1 \mathcal{P}^{2i+2,2i+3}_1 + \text{H.c.} \right)\\
	&\ +\left(\frac{1}{2}\frac{t^4}{V^3} -\left(\frac{19}{4}+\frac{103}{27}\phi^{-1}\right)\frac{t^6}{V^5}\right) \sum_i \mathcal{P}^{i,i+1}_1 + \left(-\frac{3}{2}\frac{t^2}{V}+\left( \frac{17}{24}+\frac{1}{2}\phi^{-1} \right)\frac{t^4}{V^3}+\left(\frac{65}{432}+\frac{1}{54}\phi^{-1}\right)\frac{t^6}{V^5}\right)L,
	\end{split}
\end{align}
\end{widetext}
which is identical to Eq.~\eqref{eq:Heff} in the main text. Finally, let us note that all the prefactors for the fourth and sixth-order terms have also been verified numerically.

\section{Connection to the Model in Ref.~\cite{GoldenChainAndMG}}
\label{app:connection}

\begin{figure}
\centering
\includegraphics{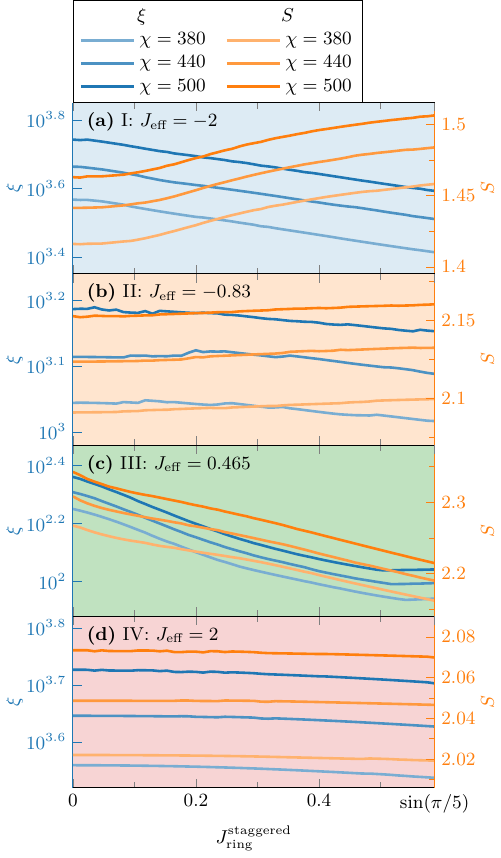}
\caption{Entanglement entropy and correlation length for $0\leq J_{\text{ring}}^{\text{staggered}} \leq \sin(\pi/5)$ and bond dimensions $\chi=380,440,500$ across all four distinct phases of the effective Hamiltonian $\mathcal{H}_{\text{eff}}$ observed in the main text, for \textbf{(a)} $J_{\text{eff}}=-2$ (phase I), \textbf{(b)} $J_{\text{eff}}=-0.83$ (phase II), \textbf{(c)} $J_{\text{eff}}=0.465$ (phase III), and \textbf{(d)} $J_{\text{eff}}=2$ (phase IV). A value of $J_{\text{ring}}^{\text{staggered}} = \sin(\pi/5)$ corresponds to $\mathcal{H}_{\text{eff}}$, while $J_{\text{ring}}^{\text{staggered}} = 0$ corresponds to the Hamiltonian $\mathcal{H}'$ studied in Ref.~\cite{GoldenChainAndMG}.}
\label{fig:ConnectPhases}
\end{figure}

\begin{figure*}
\begin{minipage}[t]{0.482\textwidth}
\includegraphics[]{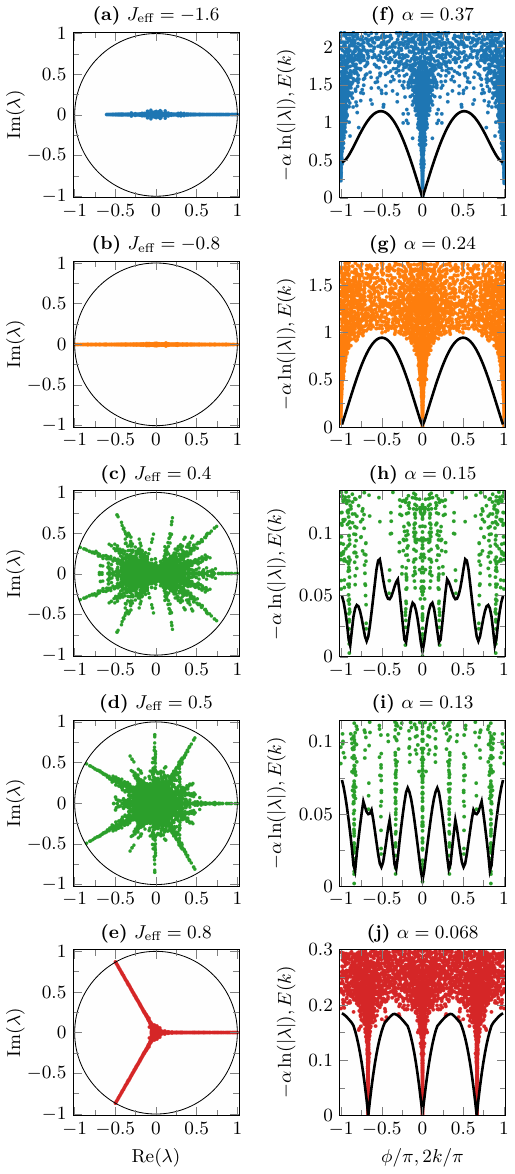}
    \caption{Transfer matrix spectra of the effective Hamiltonian $\mathcal{H}_{\text{eff}}$ obtained from iMPS simualtions using two-site unit cells for \textbf{(a)} $J_{\text{eff}}=-1.6$, \textbf{(b)} $J_{\text{eff}}=-0.8$, \textbf{(c)} $J_{\text{eff}}=0.4$, \textbf{(d)} $J_{\text{eff}}=0.4$ and \textbf{(e)} $J_{\text{eff}}=0.8$, where the bond dimension was set to $\chi=100$ in all cases. The plots \textbf{(f)} - \textbf{(j)} correspond to the same spectra as depicted in \textbf{(a)} - \textbf{(e)} but plotted with respect to the absolute values and complex phases rather than real and imaginary parts. The solid lines correspond to the lowest-lying excitation $E(k)$ for each $J_{\text{eff}}$.}
   \label{fig:TMS1}
\end{minipage}
\hfill
\begin{minipage}[t]{0.482\textwidth}
\includegraphics[]{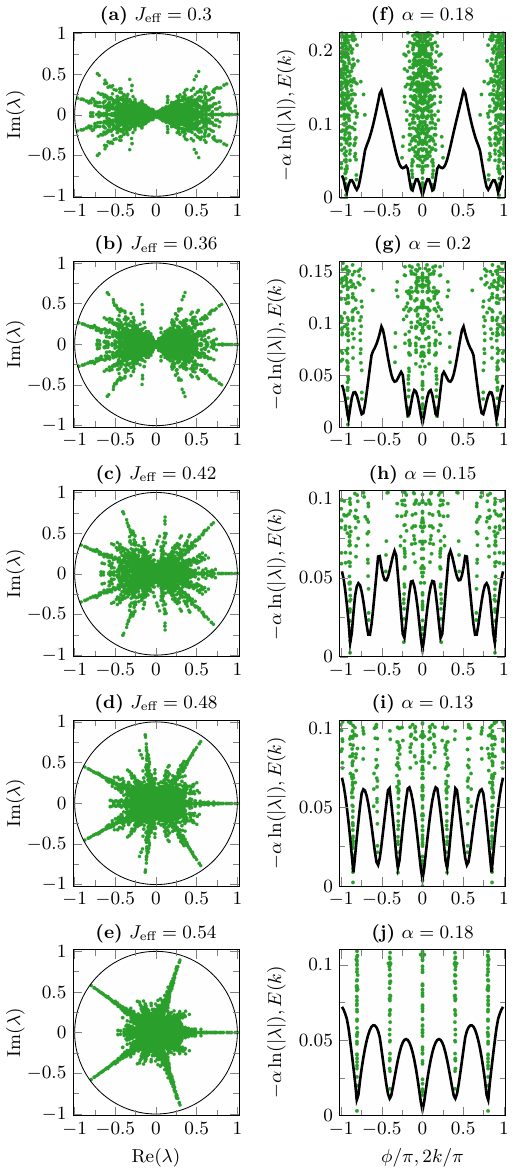}
    \caption{Transfer matrix spectra of the effective Hamiltonian $\mathcal{H}_{\text{eff}}$ obtained from iMPS simualtions using two-site unit cells for \textbf{(a)} $J_{\text{eff}}=0.3$, \textbf{(b)} $J_{\text{eff}}=0.36$, \textbf{(c)} $J_{\text{eff}}=0.42$, \textbf{(d)} $J_{\text{eff}}=0.48$ and \textbf{(e)} $J_{\text{eff}}=0.54$, where the bond dimension was set to $\chi=100$ in all cases. The plots \textbf{(f)} - \textbf{(j)} correspond to the same spectra as depicted in \textbf{(a)} - \textbf{(e)} but plotted with respect to the absolute values and complex phases rather than real and imaginary parts. The solid lines correspond to the lowest-lying excitation $E(k)$ for each $J_{\text{eff}}$. All values of $J_{\text{eff}}$ are chosen such that $\mathcal{H}_{\text{eff}}$ realizes phase III, which supports incommensurate correlations.}
   \label{fig:TMS2}
\end{minipage}%
\end{figure*}

In this section, we focus on the connection between the (rescaled) effective Hamiltonian $\mathcal{H}_{\text{eff}}$ in Eq.~\eqref{eq:Heff2} and the model studied in Ref.~\cite{GoldenChainAndMG}, which acts on the same Hilbert space as $\mathcal{H}_{\text{eff}}$, and can be written as
\begin{align}
	\widetilde{\mathcal{H}}' = -\cos\theta \sum_{i}\mathcal{P}^{i,i+1}_1 + \sin \theta \sum_i \mathcal{P}^{i,i+1,i+2}_{\tau}.
\end{align}
Here $\mathcal{P}^{i,i+1,i+2}_{\tau}$ denotes the projector of the $i$-th, $(i+1)$-th and $(i+2)$-th anyons onto their common $\tau$ fusion channel and we assume periodic boundary conditions. This projector can be rewritten in terms of $\mathcal{P}^{i,i+1}_1$ and $\mathcal{P}^{i+1,i+2}_1$ projectors as
\begin{align}
	\begin{split}
	\mathcal{P}^{i,i+1,i+2}_{\tau} = \phi \big\lbrace & \big( 1-\mathcal{P}^{i,i+1}_1 \big) \mathcal{P}^{i+1,i+2}_1 \\
	&+ \big(1-\mathcal{P}^{i+1,i+2}_1\big)\mathcal{P}^{i,i+1}_1 \big\rbrace,
	\end{split}
\end{align}
where $\phi = (1+\sqrt{5})/2$ is the golden ratio. The Hamiltonian $\widetilde{\mathcal{H}}'$ can thus be reexpressed as
\begin{align}
	\widetilde{\mathcal{H}}' = (2\phi\sin\theta  -\cos\theta) \mathcal{H}_{\text{gc}} - \phi\sin\theta \mathcal{H}_{\text{ring}}^{\text{uniform}},
\end{align}
where $\mathcal{H}_{\text{gc}}$ denotes the golden chain Hamiltonian (cf. Eq.~\eqref{eq:Heff2}) and we have split the contribution $\mathcal{H}_{\text{ring}}$ to $\mathcal{H}_{\text{eff}}$ into the translation symmetric contribution $\mathcal{H}_{\text{ring}}^{\text{uniform}}$ and the staggered contribution $\mathcal{H}_{\text{ring}}^{\text{staggered}}$ as
\begin{align}
	\mathcal{H}_{\text{ring}}^{\text{uniform}} =& \sum_{i} \big( \mathcal{P}^{i,i+1}_1\mathcal{P}^{i+1,i+2}_1 + \text{H.c.}\big), \\
	\begin{split}
	\mathcal{H}_{\text{ring}}^{\text{staggered}} =& \sum_{i} \big( -i\mathcal{P}^{2i,2i+1}_1\mathcal{P}^{2i+1,2i+2}_1 \\ &\quad \quad \hspace{2.5pt} +i\mathcal{P}^{2i+1,2i+2}_1\mathcal{P}^{2i+2,2i+3}_1 + \text{H.c.}\big). \hspace{-10pt}
	\end{split}
\end{align}
With that, the effective Hamiltonian $\mathcal{H}_{\text{eff}}$ in Eq.~\eqref{eq:Heff2} takes the form
\begin{align}
		\mathcal{H}_{\text{eff}} =& J_{\text{eff}}\mathcal{H}_{\text{gc}} + \cos(\pi/5) \mathcal{H}_{\text{ring}}^{\text{uniform}} +J_{\text{ring}}^{\text{staggered}} \mathcal{H}_{\text{ring}}^{\text{staggered}}
\end{align}
with $J_{\text{ring}}^{\text{staggered}}=\sin(\pi/5)$, and after rescaling $\widetilde{\mathcal{H}}'$ such that we obtain\footnote{Note that unlike in Ref.~\cite{GoldenChainAndMG}, the sign of $\mathcal{H}_{\text{ring}}^{\text{uniform}}$ arising from the perturbation theory in App.~\ref{App:PT} is necessarily positive, and we will thus focus on this case.}
\begin{align}
	\mathcal{H}' = \cos\left(\pi/5\right) \left(\phi^{-1}\cot\theta  -2\right) \mathcal{H}_{\text{gc}}  + \cos(\pi/5) \mathcal{H}_{\text{ring}}^{\text{uniform}},
\end{align}
it can be seen that the main difference between the Hamiltonian in Ref.~\cite{GoldenChainAndMG} and the effective Hamiltonian $\mathcal{H}_{\text{eff}}$ is the staggered term $\mathcal{H}_{\text{ring}}^{\text{staggered}}$. This connection naturally raises the question whether or not the phases discussed in the main text (cf. Fig.~\ref{fig:phase_diagram}) are related to the ones observed in Ref.~\cite{GoldenChainAndMG}.

In both the effective Hamiltonian $\mathcal{H}_{\text{eff}}$ and the model in Ref.~\cite{GoldenChainAndMG}, gapless phases described by the tricritical Ising CFT~\cite{FRIEDAN198537, francesco2012conformal} and the three-state Potts CFT~\cite{DOTSENKO198454, francesco2012conformal} can be found. These phases must be connected as we can take the limit $J_{\text{eff}}\rightarrow \pm \infty$ and then tune $J_{\text{ring}}^{\text{staggered}}=\sin(\pi/5) \rightarrow 0$. This process cannot drive a phase transition since we have already observed in the main text that these two phases are stable against adding the $\mathcal{H}_{\text{ring}}^{\text{staggered}}$ term. The phases are therefore adiabatically connected. This can be confirmed numerically by monitoring the entanglement entropy and correlation length for different bond dimensions while tuning $J_{\text{ring}}^{\text{staggered}}$ from $\sin(\pi/5)$ to $0$. The results associated with the tricritial Ising CFT phase (phase I) and the three-state Potts CFT phase (phase IV) are shown in Fig.~\ref{fig:ConnectPhases}\textbf{(a)} and \textbf{(d)}, where $J_{\text{eff}}$ is fixed to $J_{\text{eff}}=-2$ and $J_{\text{eff}}=2$, respectively. Due to the absence of any singularities in the entanglement entropy and the correlation length, the numerics suggest that there exists an adiabatic path between phase I of the effective Hamiltonian $\mathcal{H}_{\text{eff}}$ and the corresponding phase of $\mathcal{H}'$ at $J_{\text{eff}}=-2$. Similarly, phase IV of $\mathcal{H}_{\text{eff}}$ appears to be adiabatically connected to the three-state Potts CFT phase of $\mathcal{H}'$ via the path described by $J_{\text{eff}}=2$. The numerical results thus confirm our expectation of phases I and IV of the effective Hamiltonian $\mathcal{H}_{\text{eff}}$ extending to the Hamiltonian $\mathcal{H}'$ studied in Ref.~\cite{GoldenChainAndMG}.

The results in Fig.~\ref{fig:ConnectPhases}\textbf{(b)} and \textbf{(c)} further suggest that phase II of $\mathcal{H}_{\text{eff}}$ corresponds to the $\mathbb{Z}_4$ phase\footnote{Note that we refer to phase II as a $\mathbb{Z}_2$ phase due to $\mathcal{H}_{\text{eff}}$ having a two-site unit cell. Upon removing the staggered contribution, i.e., for $\mathcal{H}'$, we have a single-site unit cell and the phase is refered to as a $\mathbb{Z}_4$ phase. The difference in naming is thus merely an effect of different unit cell sizes.} observed in Ref.~\cite{GoldenChainAndMG}, and that the phases with incommensurate correlations of $\mathcal{H}_{\text{eff}}$ and $\mathcal{H}'$ are adiabatically connected. It therefore follows that all the phases of $\mathcal{H}_{\text{eff}}$ observed in the main text correspond to phases of the Hamiltonian $\mathcal{H}'$. Note however that in Ref.~\cite{GoldenChainAndMG}, a gapped phase that does not have a counterpart in the phase diagram of the effective Hamiltonian $\mathcal{H}_{\text{eff}}$ was found.

We can thus conclude that adding the staggered contribution $\mathcal{H}_{\text{ring}}^{\text{staggered}}$ (with coupling constant as specified above) to $\mathcal{H}'$ does not only shift the phase transition points, but does also annihilate the gapped phase observed in Ref.~\cite{GoldenChainAndMG}.

\section{Additional Transfer Matrix Spectra}
\label{App:TMS}

In this section, additional numerical data for the transfer matrix spectra discussed in Sec.~\ref{sec:TMS} are provided in Figs.~\ref{fig:TMS1} and \ref{fig:TMS2}, which show the complex eigenvalues $\lambda$ of the transfer matrices both in terms of their real and imaginary parts (\textbf{a} - \textbf{e}) and in terms of their magnitudes $-\ln(|\lambda|)$ and complex phases $\phi=\arg(\lambda)$ (\textbf{f} - \textbf{j}). The latter plots also show the dispersion of the lowest-lying excitation as solid lines. Note that the values of $J_{\text{eff}}$ selected for phases I, II and IV correspond to the respective values in Fig.~\ref{fig:TMS_phases}. All remaining values of $J_{\text{eff}}$ are chosen such that the system realizes phase III. The incommensurate nature of this phase can be confirmed when comparing the transfer matrix spectra in Figs.~\ref{fig:TMS1} and \ref{fig:TMS2} across the different values of $J_{\text{eff}}$.

\bibliography{./references.bib}

@dataset{zenodo,
author = {Nico Kirchner and Roderich Moessner and Adam Gammon-Smith and Frank Pollmann},
title = "{Phases of Interacting Fibonacci Anyons on a Ladder at Half-Filling}",
month = {7},
year = {2025},
publisher = {Zenodo},
doi = {10.5281/zenodo.16563758},
url = {https://doi.org/10.5281/zenodo.16563758},
note = "{Zenodo}",
}

@article{hubbard,
  title={Electron correlations in narrow energy bands},
  author={Hubbard, John},
  journal={Proc. R. Soc. Lond. A},
  volume={276},
  number={1365},
  pages={238--257},
  year={1963},
  publisher={The Royal Society London},
  doi = {https://doi.org/10.1098/rspa.1963.0204},
}

@article{hubbard2,
  title = "{Effect of Correlation on the Ferromagnetism of Transition Metals}",
  author = {Gutzwiller, Martin C.},
  journal = {Phys. Rev. Lett.},
  volume = {10},
  issue = {5},
  pages = {159--162},
  numpages = {0},
  year = {1963},
  month = {Mar},
  publisher = {American Physical Society},
  doi = {10.1103/PhysRevLett.10.159},
  url = {https://link.aps.org/doi/10.1103/PhysRevLett.10.159}
}

@article{hubbard3,
    author = {Kanamori, Junjiro},
    title = "{Electron Correlation and Ferromagnetism of Transition Metals}",
    journal = {Progress of Theoretical Physics},
    volume = {30},
    number = {3},
    pages = {275-289},
    year = {1963},
    month = {09},
    issn = {0033-068X},
    doi = {10.1143/PTP.30.275},
    url = {https://doi.org/10.1143/PTP.30.275},
}

@article{hubbard_phases,
   author = "Qin, Mingpu and Schäfer, Thomas and Andergassen, Sabine and Corboz, Philippe and Gull, Emanuel",
   title = "{The Hubbard Model: A Computational Perspective}", 
   journal= "Annual Review of Condensed Matter Physics",
   year = "2022",
   volume = "13",
   number = "Volume 13, 2022",
   pages = "275-302",
   doi = "https://doi.org/10.1146/annurev-conmatphys-090921-033948",
   url = "https://www.annualreviews.org/content/journals/10.1146/annurev-conmatphys-090921-033948",
   publisher = "Annual Reviews",
   issn = "1947-5462",
   type = "Journal Article",
  }

@article{Leinaas1977,
author={Leinaas, J. M.
and Myrheim, J.},
title={On the theory of identical particles},
journal={Il Nuovo Cimento B (1971-1996)},
year={1977},
month={Jan},
day={01},
volume={37},
number={1},
pages={1-23},
issn={1826-9877},
doi={10.1007/BF02727953},
url={https://doi.org/10.1007/BF02727953}
}

@article{PhysRevLett.48.1144,
  title = "{Magnetic Flux, Angular Momentum, and Statistics}",
  author = {Wilczek, Frank},
  journal = {Phys. Rev. Lett.},
  volume = {48},
  issue = {17},
  pages = {1144--1146},
  numpages = {0},
  year = {1982},
  month = {Apr},
  publisher = {American Physical Society},
  doi = {10.1103/PhysRevLett.48.1144},
  url = {https://link.aps.org/doi/10.1103/PhysRevLett.48.1144}
}

@article{PhysRevLett.49.957,
  title = "{Quantum Mechanics of Fractional-Spin Particles}",
  author = {Wilczek, Frank},
  journal = {Phys. Rev. Lett.},
  volume = {49},
  issue = {14},
  pages = {957--959},
  numpages = {0},
  year = {1982},
  month = {Oct},
  publisher = {American Physical Society},
  doi = {10.1103/PhysRevLett.49.957},
  url = {https://link.aps.org/doi/10.1103/PhysRevLett.49.957}
}

@article{QSL0,
  title = "{Resonating Valence Bond Phase in the Triangular Lattice Quantum Dimer Model}",
  author = {Moessner, R. and Sondhi, S. L.},
  journal = {Phys. Rev. Lett.},
  volume = {86},
  issue = {9},
  pages = {1881--1884},
  numpages = {0},
  year = {2001},
  month = {Feb},
  publisher = {American Physical Society},
  doi = {10.1103/PhysRevLett.86.1881},
  url = {https://link.aps.org/doi/10.1103/PhysRevLett.86.1881}
}

@article{QSL1,
  title = "{Fractionalization in an easy-axis Kagome antiferromagnet}",
  author = {Balents, L. and Fisher, M. P. A. and Girvin, S. M.},
  journal = {Phys. Rev. B},
  volume = {65},
  issue = {22},
  pages = {224412},
  numpages = {8},
  year = {2002},
  month = {May},
  publisher = {American Physical Society},
  doi = {10.1103/PhysRevB.65.224412},
  url = {https://link.aps.org/doi/10.1103/PhysRevB.65.224412}
}

@article{QSL3,
  title = "{Pyrochlore photons: The $U(1)$ spin liquid in a $S=\frac{1}{2}$ three-dimensional frustrated magnet}",
  author = {Hermele, Michael and Fisher, Matthew P. A. and Balents, Leon},
  journal = {Phys. Rev. B},
  volume = {69},
  issue = {6},
  pages = {064404},
  numpages = {21},
  year = {2004},
  month = {Feb},
  publisher = {American Physical Society},
  doi = {10.1103/PhysRevB.69.064404},
  url = {https://link.aps.org/doi/10.1103/PhysRevB.69.064404}
}

@article{QSL4,
  title = "{String-net condensation: A physical mechanism for topological phases}",
  author = {Levin, Michael A. and Wen, Xiao-Gang},
  journal = {Phys. Rev. B},
  volume = {71},
  issue = {4},
  pages = {045110},
  numpages = {21},
  year = {2005},
  month = {Jan},
  publisher = {American Physical Society},
  doi = {10.1103/PhysRevB.71.045110},
  url = {https://link.aps.org/doi/10.1103/PhysRevB.71.045110}
}

@article{QSL5,
	author = {Kitaev, Alexei},
	doi = {10.1016/j.aop.2005.10.005},
	journal = {Ann. Phys.},
	number = {1},
	pages = {2--111},
	title = {Anyons in an exactly solved model and beyond},
	url = {http://dx.doi.org/10.1016/j.aop.2005.10.005},
	volume = {321},
	year = {2006}
}

@article{QSL6,
doi = {10.1088/0034-4885/80/1/016502},
url = {https://dx.doi.org/10.1088/0034-4885/80/1/016502},
year = {2017},
month = {nov},
publisher = {IOP Publishing},
volume = {80},
number = {1},
pages = {016502},
author = {Savary, Lucile and Balents, Leon},
title = {Quantum spin liquids: a review},
journal = {Rep. Prog. Phys.},
}

@article{TQC1,
  title={A modular functor which is universal for quantum computation},
  author={Freedman, Michael H and Larsen, Michael and Wang, Zhenghan},
  journal={Commun. Math. Phys.},
  volume={227},
  number={3},
  pages={605--622},
  year={2002},
  publisher={Springer},
  doi = {https://doi.org/10.1007/s002200200645}
}

@article{TQC2,
title = {Fault-tolerant quantum computation by anyons},
journal = {Annals of Physics},
volume = {303},
number = {1},
pages = {2-30},
year = {2003},
issn = {0003-4916},
doi = {https://doi.org/10.1016/S0003-4916(02)00018-0},
url = {https://www.sciencedirect.com/science/article/pii/S0003491602000180},
author = {A.Yu. Kitaev},
}

@article{TQC3,
  title = "{Non-Abelian anyons and topological quantum computation}",
  author = {Nayak, Chetan and Simon, Steven H. and Stern, Ady and Freedman, Michael and Das Sarma, Sankar},
  journal = {Rev. Mod. Phys.},
  volume = {80},
  issue = {3},
  pages = {1083--1159},
  numpages = {0},
  year = {2008},
  month = {Sep},
  publisher = {American Physical Society},
  doi = {10.1103/RevModPhys.80.1083},
  url = {https://link.aps.org/doi/10.1103/RevModPhys.80.1083}
}

@article{TQC4,
  title = "{Universal Topological Quantum Computation from a Superconductor-Abelian Quantum Hall Heterostructure}",
  author = {Mong, Roger S. K. and Clarke, David J. and Alicea, Jason and Lindner, Netanel H. and Fendley, Paul and Nayak, Chetan and Oreg, Yuval and Stern, Ady and Berg, Erez and Shtengel, Kirill and Fisher, Matthew P. A.},
  journal = {Phys. Rev. X},
  volume = {4},
  issue = {1},
  pages = {011036},
  numpages = {41},
  year = {2014},
  month = {Mar},
  publisher = {American Physical Society},
  doi = {10.1103/PhysRevX.4.011036},
  url = {https://link.aps.org/doi/10.1103/PhysRevX.4.011036}
}

@article{TQC5,
doi = {10.1088/2058-9565/aacad2},
url = {https://dx.doi.org/10.1088/2058-9565/aacad2},
year = {2018},
month = {jul},
publisher = {IOP Publishing},
volume = {3},
number = {4},
pages = {045004},
author = {Field, Bernard and Simula, Tapio},
title = "{Introduction to topological quantum computation with non-Abelian anyons}",
journal = {Quantum Sci. Technol.},
}

@article{FQH1,
  title = "{Anomalous Quantum Hall Effect: An Incompressible Quantum Fluid with Fractionally Charged Excitations}",
  author = {Laughlin, R. B.},
  journal = {Phys. Rev. Lett.},
  volume = {50},
  issue = {18},
  pages = {1395--1398},
  numpages = {0},
  year = {1983},
  month = {May},
  publisher = {American Physical Society},
  doi = {10.1103/PhysRevLett.50.1395},
  url = {https://link.aps.org/doi/10.1103/PhysRevLett.50.1395}
}

@article{FQH2,
  title = "{Statistics of Quasiparticles and the Hierarchy of Fractional Quantized Hall States}",
  author = {Halperin, B. I.},
  journal = {Phys. Rev. Lett.},
  volume = {52},
  issue = {18},
  pages = {1583--1586},
  numpages = {0},
  year = {1984},
  month = {Apr},
  publisher = {American Physical Society},
  doi = {10.1103/PhysRevLett.52.1583},
  url = {https://link.aps.org/doi/10.1103/PhysRevLett.52.1583}
}

@article{FQH3,
  title = "{Fractional Statistics and the Quantum Hall Effect}",
  author = {Arovas, Daniel and Schrieffer, J. R. and Wilczek, Frank},
  journal = {Phys. Rev. Lett.},
  volume = {53},
  issue = {7},
  pages = {722--723},
  numpages = {0},
  year = {1984},
  month = {Aug},
  publisher = {American Physical Society},
  doi = {10.1103/PhysRevLett.53.722},
  url = {https://link.aps.org/doi/10.1103/PhysRevLett.53.722}
}

@article{FQH4,
  title = "{Ground-state degeneracy of the fractional quantum Hall states in the presence of a random potential and on high-genus Riemann surfaces}",
  author = {Wen, X. G. and Niu, Q.},
  journal = {Phys. Rev. B},
  volume = {41},
  issue = {13},
  pages = {9377--9396},
  numpages = {0},
  year = {1990},
  month = {May},
  publisher = {American Physical Society},
  doi = {10.1103/PhysRevB.41.9377},
  url = {https://link.aps.org/doi/10.1103/PhysRevB.41.9377}
}

@article{FQH5,
title = "{Anyons and the quantum Hall effect—A pedagogical review}",
journal = {Annals of Physics},
volume = {323},
number = {1},
pages = {204-249},
year = {2008},
issn = {0003-4916},
doi = {https://doi.org/10.1016/j.aop.2007.10.008},
url = {https://www.sciencedirect.com/science/article/pii/S0003491607001674},
author = {Ady Stern},
}

@article{FQH6,
  title = "{Beyond paired quantum Hall states: Parafermions and incompressible states in the first excited Landau level}",
  author = {Read, N. and Rezayi, E.},
  journal = {Phys. Rev. B},
  volume = {59},
  issue = {12},
  pages = {8084--8092},
  numpages = {0},
  year = {1999},
  month = {Mar},
  publisher = {American Physical Society},
  doi = {10.1103/PhysRevB.59.8084},
  url = {https://link.aps.org/doi/10.1103/PhysRevB.59.8084}
}

@article{PerturbationTheory,
doi = {10.1088/0022-3719/2/12/301},
url = {https://dx.doi.org/10.1088/0022-3719/2/12/301},
year = {1969},
month = {dec},
publisher = {},
volume = {2},
number = {12},
pages = {2161},
author = {C E Soliverez},
title = {An effective {H}amiltonian and time-independent perturbation theory},
journal = {J. Phys. C: Solid State Phys.},
}

@article{GoldenChain,
  title = "{Interacting Anyons in Topological Quantum Liquids: The Golden Chain}",
  author = {Feiguin, Adrian and Trebst, Simon and Ludwig, Andreas W. W. and Troyer, Matthias and Kitaev, Alexei and Wang, Zhenghan and Freedman, Michael H.},
  journal = {Phys. Rev. Lett.},
  volume = {98},
  issue = {16},
  pages = {160409},
  numpages = {4},
  year = {2007},
  month = {Apr},
  publisher = {American Physical Society},
  doi = {10.1103/PhysRevLett.98.160409},
  url = {https://link.aps.org/doi/10.1103/PhysRevLett.98.160409}
}

@article{anyonic_defects,
author={Buican, Matthew
and Gromov, Andrey},
title="{Anyonic Chains, Topological Defects, and Conformal Field Theory}",
journal={Commun. Math. Phys.},
year={2017},
month={Dec},
day={01},
volume={356},
number={3},
pages={1017-1056},
issn={1432-0916},
doi={10.1007/s00220-017-2995-6},
url={https://doi.org/10.1007/s00220-017-2995-6}
}

@article{PhysRevB.90.075129,
  title = "{Topological insulating phases of non-Abelian anyonic chains}",
  author = {DeGottardi, Wade},
  journal = {Phys. Rev. B},
  volume = {90},
  issue = {7},
  pages = {075129},
  numpages = {8},
  year = {2014},
  month = {Aug},
  publisher = {American Physical Society},
  doi = {10.1103/PhysRevB.90.075129},
  url = {https://link.aps.org/doi/10.1103/PhysRevB.90.075129}
}

@article{IntroFibAndGC,
    author = {Trebst, Simon and Troyer, Matthias and Wang, Zhenghan and Ludwig, Andreas W. W.},
    title = "{A Short Introduction to Fibonacci Anyon Models}",
    journal = {Progress of Theoretical Physics Supplement},
    volume = {176},
    pages = {384-407},
    year = {2008},
    month = {06},
    issn = {0375-9687},
    doi = {10.1143/PTPS.176.384},
    url = {https://doi.org/10.1143/PTPS.176.384},
}

@article{GoldenLadder,
  title = "{Quantum spin ladders of non-Abelian anyons}",
  author = {Poilblanc, Didier and Ludwig, Andreas W. W. and Trebst, Simon and Troyer, Matthias},
  journal = {Phys. Rev. B},
  volume = {83},
  issue = {13},
  pages = {134439},
  numpages = {12},
  year = {2011},
  month = {Apr},
  publisher = {American Physical Society},
  doi = {10.1103/PhysRevB.83.134439},
  url = {https://link.aps.org/doi/10.1103/PhysRevB.83.134439}
}

@article{YangLeeAnyons,
doi = {10.1088/1367-2630/13/4/045006},
url = {https://dx.doi.org/10.1088/1367-2630/13/4/045006},
year = {2011},
month = {apr},
publisher = {},
volume = {13},
number = {4},
pages = {045006},
author = {E Ardonne and J Gukelberger and A W W Ludwig and S Trebst and M Troyer},
title = "{Microscopic models of interacting Yang–Lee anyons}",
journal = {New J. Phys.},
}

@article{AnyonLadderAyeni,
  title = "{Phase transitions on a ladder of braided non-Abelian anyons}",
  author = {Ayeni, Babatunde M. and Pfeifer, Robert N. C. and Brennen, Gavin K.},
  journal = {Phys. Rev. B},
  volume = {98},
  issue = {4},
  pages = {045432},
  numpages = {14},
  year = {2018},
  month = {Jul},
  publisher = {American Physical Society},
  doi = {10.1103/PhysRevB.98.045432},
  url = {https://link.aps.org/doi/10.1103/PhysRevB.98.045432}
}

@article{PhysRevB.93.165128,
  title = {Simulation of braiding anyons using matrix product states},
  author = {Ayeni, Babatunde M. and Singh, Sukhwinder and Pfeifer, Robert N. C. and Brennen, Gavin K.},
  journal = {Phys. Rev. B},
  volume = {93},
  issue = {16},
  pages = {165128},
  numpages = {18},
  year = {2016},
  month = {Apr},
  publisher = {American Physical Society},
  doi = {10.1103/PhysRevB.93.165128},
  url = {https://link.aps.org/doi/10.1103/PhysRevB.93.165128}
}

@article{GoldenChainAndMG,
  title = "{Collective States of Interacting Fibonacci Anyons}",
  author = {Trebst, Simon and Ardonne, Eddy and Feiguin, Adrian and Huse, David A. and Ludwig, Andreas W. W. and Troyer, Matthias},
  journal = {Phys. Rev. Lett.},
  volume = {101},
  issue = {5},
  pages = {050401},
  numpages = {4},
  year = {2008},
  month = {Jul},
  publisher = {American Physical Society},
  doi = {10.1103/PhysRevLett.101.050401},
  url = {https://link.aps.org/doi/10.1103/PhysRevLett.101.050401}
}

@article{PhysRevLett.108.207201,
  title = "{Fractionalization of Itinerant Anyons in One-Dimensional Chains}",
  author = {Poilblanc, Didier and Troyer, Matthias and Ardonne, Eddy and Bonderson, Parsa},
  journal = {Phys. Rev. Lett.},
  volume = {108},
  issue = {20},
  pages = {207201},
  numpages = {5},
  year = {2012},
  month = {May},
  publisher = {American Physical Society},
  doi = {10.1103/PhysRevLett.108.207201},
  url = {https://link.aps.org/doi/10.1103/PhysRevLett.108.207201}
}

@article{PhysRevB.87.085106,
  title = "{One-dimensional itinerant interacting non-Abelian anyons}",
  author = {Poilblanc, Didier and Feiguin, Adrian and Troyer, Matthias and Ardonne, Eddy and Bonderson, Parsa},
  journal = {Phys. Rev. B},
  volume = {87},
  issue = {8},
  pages = {085106},
  numpages = {18},
  year = {2013},
  month = {Feb},
  publisher = {American Physical Society},
  doi = {10.1103/PhysRevB.87.085106},
  url = {https://link.aps.org/doi/10.1103/PhysRevB.87.085106}
}

@article{AnyonSpin1Chains,
  title = "{Anyonic quantum spin chains: Spin-1 generalizations and topological stability}",
  author = {Gils, C. and Ardonne, E. and Trebst, S. and Huse, D. A. and Ludwig, A. W. W. and Troyer, M. and Wang, Z.},
  journal = {Phys. Rev. B},
  volume = {87},
  issue = {23},
  pages = {235120},
  numpages = {33},
  year = {2013},
  month = {Jun},
  publisher = {American Physical Society},
  doi = {10.1103/PhysRevB.87.235120},
  url = {https://link.aps.org/doi/10.1103/PhysRevB.87.235120}
}

@article{PhysRevLett.126.163201,
  title = "{Bosonic Continuum Theory of One-Dimensional Lattice Anyons}",
  author = {Bonkhoff, Martin and J\"agering, Kevin and Eggert, Sebastian and Pelster, Axel and Thorwart, Michael and Posske, Thore},
  journal = {Phys. Rev. Lett.},
  volume = {126},
  issue = {16},
  pages = {163201},
  numpages = {7},
  year = {2021},
  month = {Apr},
  publisher = {American Physical Society},
  doi = {10.1103/PhysRevLett.126.163201},
  url = {https://link.aps.org/doi/10.1103/PhysRevLett.126.163201}
}

@article{PhysRevB.108.155134,
  title = "{Coherence properties of the repulsive anyon-Hubbard dimer}",
  author = {Bonkhoff, Martin and J\"ager, Simon B. and Schneider, Imke and Pelster, Axel and Eggert, Sebastian},
  journal = {Phys. Rev. B},
  volume = {108},
  issue = {15},
  pages = {155134},
  numpages = {8},
  year = {2023},
  month = {Oct},
  publisher = {American Physical Society},
  doi = {10.1103/PhysRevB.108.155134},
  url = {https://link.aps.org/doi/10.1103/PhysRevB.108.155134}
}

@misc{bonkhoff2024anyonicphasetransitions1d,
      title="{Anyonic phase transitions in the 1D extended Hubbard model with fractional statistics}", 
      author={Martin Bonkhoff and Kevin Jägering and Shijie Hu and Axel Pelster and Sebastian Eggert and Imke Schneider},
      year={2024},
      eprint={2410.00089},
      archivePrefix={arXiv},
      primaryClass={cond-mat.str-el},
}

@article{PhysRevB.43.2661,
  title = {Braid group and anyons on a cylinder},
  author = {Hatsugai, Yasuhiro and Kohmoto, Mahito and Wu, Yong-Shi},
  journal = {Phys. Rev. B},
  volume = {43},
  issue = {4},
  pages = {2661--2677},
  numpages = {0},
  year = {1991},
  month = {Feb},
  publisher = {American Physical Society},
  doi = {10.1103/PhysRevB.43.2661},
  url = {https://link.aps.org/doi/10.1103/PhysRevB.43.2661}
}

@article{PhysRevB.43.10761,
  title = "{Anyons on a torus: Braid group, Aharonov-Bohm period, and numerical study}",
  author = {Hatsugai, Yasuhiro and Kohmoto, Mahito and Wu, Yong-Shi},
  journal = {Phys. Rev. B},
  volume = {43},
  issue = {13},
  pages = {10761--10768},
  numpages = {0},
  year = {1991},
  month = {May},
  publisher = {American Physical Society},
  doi = {10.1103/PhysRevB.43.10761},
  url = {https://link.aps.org/doi/10.1103/PhysRevB.43.10761}
}

@article{PhysRevB.48.13742,
  title = {Flux quantization for semions on a torus},
  author = {Kallin, C.},
  journal = {Phys. Rev. B},
  volume = {48},
  issue = {18},
  pages = {13742--13748},
  numpages = {0},
  year = {1993},
  month = {Nov},
  publisher = {American Physical Society},
  doi = {10.1103/PhysRevB.48.13742},
  url = {https://link.aps.org/doi/10.1103/PhysRevB.48.13742}
}

@article{PhysRevB.107.195129,
  title = "{Numerical simulation of non-Abelian anyons}",
  author = {Kirchner, Nico and Millar, Darragh and Ayeni, Babatunde M. and Smith, Adam and Slingerland, Joost K. and Pollmann, Frank},
  journal = {Phys. Rev. B},
  volume = {107},
  issue = {19},
  pages = {195129},
  numpages = {33},
  year = {2023},
  month = {May},
  publisher = {American Physical Society},
  doi = {10.1103/PhysRevB.107.195129},
  url = {https://link.aps.org/doi/10.1103/PhysRevB.107.195129}
}

@phdthesis{darragh_thesis,
title="{Interacting Anyons in One and Two Dimensions: Strong Zero Modes in Anyon Chains and Non-Abelian Anyons on a Torus}",
author={Darragh Millar},
year={2021},
url = {https://mural.maynoothuniversity.ie/14865/},
school = {National University of Ireland Maynooth},
}

@mastersthesis{nico_master,
  author  = {Nico Kirchner},
  title   = "{Simulation of Anyon Dynamics}",
  school  = {Technische Universit{\"a}t M{\"u}nchen},
  year    = {2021},
  doi     = {10.5281/zenodo.7566432},
}

@book{topologicalquantum,
    author = {Simon, Steven H.},
    title = "{Topological Quantum}",
    publisher = {Oxford University Press},
    year = {2023},
    month = {09},
    isbn = {9780198886723},
    doi = {10.1093/oso/9780198886723.001.0001},
    url = {https://doi.org/10.1093/oso/9780198886723.001.0001},
}

@article{1506.05805,
  title={Topological superconductors and category theory},
  author={Bernevig, Andrei and Neupert, Titus},
  journal={Lecture Notes of the Les Houches Summer School: Topological Aspects of Condensed Matter Physics},
  publisher={Oxford University Press},
  pages={63--121},
  year={2017}
}

@article{0707.4206,
title = "{Interferometry of non-Abelian anyons}",
journal = {Annals of Physics},
volume = {323},
number = {11},
pages = {2709-2755},
year = {2008},
issn = {0003-4916},
doi = {https://doi.org/10.1016/j.aop.2008.01.012},
url = {https://www.sciencedirect.com/science/article/pii/S0003491608000171},
author = {Parsa Bonderson and Kirill Shtengel and J.K. Slingerland},
}

@phdthesis{bonderson_2007,
title="{Non-Abelian anyons and interferometry}",
author={Bonderson, Parsa Hassan},
year={2007},
school={California Institute of Technology},
}

@article{2102.05677,
Author = {Parsa Bonderson},
Title = "{Measuring Topological Order}",
Year = {2021},
Howpublished = {Phys. Rev. Research 3, 033110 (2021)},
Doi = {10.1103/PhysRevResearch.3.033110},
journal = {Phys. Rev. Research},
month = {aug},
volume = {3},
number = {3},
pages = {033110},
}

@article{InsideOutsideBases,
title = {Anyonic entanglement and topological entanglement entropy},
journal = {Ann. of Phys.},
volume = {385},
pages = {399-468},
year = {2017},
issn = {0003-4916},
doi = {https://doi.org/10.1016/j.aop.2017.07.018},
url = {https://www.sciencedirect.com/science/article/pii/S0003491617302178},
author = {Parsa Bonderson and Christina Knapp and Kaushal Patel},
}

@misc{kong2022invitationtopologicalorderscategory,
      title={An invitation to topological orders and category theory}, 
      author={Liang Kong and Zhi-Hao Zhang},
      year={2022},
      eprint={2205.05565},
      archivePrefix={arXiv},
      primaryClass={cond-mat.str-el},
}

@article{PhysRevLett.103.110403,
  title = "{Splitting the Topological Degeneracy of Non-Abelian Anyons}",
  author = {Bonderson, Parsa},
  journal = {Phys. Rev. Lett.},
  volume = {103},
  issue = {11},
  pages = {110403},
  numpages = {4},
  year = {2009},
  month = {Sep},
  publisher = {American Physical Society},
  doi = {10.1103/PhysRevLett.103.110403},
  url = {https://link.aps.org/doi/10.1103/PhysRevLett.103.110403}
}

@article{PhysRevB.95.115136,
  title = "{Fibonacci anyons and charge density order in the 12/5 and 13/5 quantum Hall plateaus}",
  author = {Mong, Roger S. K. and Zaletel, Michael P. and Pollmann, Frank and Papi\ifmmode \acute{c}\else \'{c}\fi{}, Zlatko},
  journal = {Phys. Rev. B},
  volume = {95},
  issue = {11},
  pages = {115136},
  numpages = {14},
  year = {2017},
  month = {Mar},
  publisher = {American Physical Society},
  doi = {10.1103/PhysRevB.95.115136},
  url = {https://link.aps.org/doi/10.1103/PhysRevB.95.115136}
}

@article{Pasquale_Calabrese_2004,
doi = {10.1088/1742-5468/2004/06/P06002},
url = {https://dx.doi.org/10.1088/1742-5468/2004/06/P06002},
year = {2004},
month = {jun},
publisher = {},
volume = {2004},
number = {06},
pages = {P06002},
author = {Pasquale Calabrese and John Cardy},
title = {Entanglement entropy and quantum field theory},
journal = {J. Stat. Mech.},
}

@article{PhysRevB.78.024410,
  title = {Scaling of entanglement support for matrix product states},
  author = {Tagliacozzo, L. and de Oliveira, Thiago. R. and Iblisdir, S. and Latorre, J. I.},
  journal = {Phys. Rev. B},
  volume = {78},
  issue = {2},
  pages = {024410},
  numpages = {14},
  year = {2008},
  month = {Jul},
  publisher = {American Physical Society},
  doi = {10.1103/PhysRevB.78.024410},
  url = {https://link.aps.org/doi/10.1103/PhysRevB.78.024410}
}

@article{PhysRevB.82.115126,
  title = {Simulation of anyons with tensor network algorithms},
  author = {Pfeifer, R. N. C. and Corboz, P. and Buerschaper, O. and Aguado, M. and Troyer, M. and Vidal, G.},
  journal = {Phys. Rev. B},
  volume = {82},
  issue = {11},
  pages = {115126},
  numpages = {16},
  year = {2010},
  month = {Sep},
  publisher = {American Physical Society},
  doi = {10.1103/PhysRevB.82.115126},
  url = {https://link.aps.org/doi/10.1103/PhysRevB.82.115126}
}

@article{PhysRevB.89.075112,
  title = {Matrix product states for anyonic systems and efficient simulation of dynamics},
  author = {Singh, Sukhwinder and Pfeifer, Robert N. C. and Vidal, Guifre and Brennen, Gavin K.},
  journal = {Phys. Rev. B},
  volume = {89},
  issue = {7},
  pages = {075112},
  numpages = {16},
  year = {2014},
  month = {Feb},
  publisher = {American Physical Society},
  doi = {10.1103/PhysRevB.89.075112},
  url = {https://link.aps.org/doi/10.1103/PhysRevB.89.075112}
}

@article{PhysRevB.92.115135,
  title = {Finite density matrix renormalization group algorithm for anyonic systems},
  author = {Pfeifer, Robert N. C. and Singh, Sukhwinder},
  journal = {Phys. Rev. B},
  volume = {92},
  issue = {11},
  pages = {115135},
  numpages = {23},
  year = {2015},
  month = {Sep},
  publisher = {American Physical Society},
  doi = {10.1103/PhysRevB.92.115135},
  url = {https://link.aps.org/doi/10.1103/PhysRevB.92.115135}
}

@article{PhysRevB.93.035124,
  title = "{Effective models of doped quantum ladders of non-Abelian anyons}",
  author = {Soni, Medha and Troyer, Matthias and Poilblanc, Didier},
  journal = {Phys. Rev. B},
  volume = {93},
  issue = {3},
  pages = {035124},
  numpages = {24},
  year = {2016},
  month = {Jan},
  publisher = {American Physical Society},
  doi = {10.1103/PhysRevB.93.035124},
  url = {https://link.aps.org/doi/10.1103/PhysRevB.93.035124}
}

@article{PhysRevLett.103.070401,
  title = "{Collective States of Interacting Anyons, Edge States, and the Nucleation of Topological Liquids}",
  author = {Gils, Charlotte and Ardonne, Eddy and Trebst, Simon and Ludwig, Andreas W. W. and Troyer, Matthias and Wang, Zhenghan},
  journal = {Phys. Rev. Lett.},
  volume = {103},
  issue = {7},
  pages = {070401},
  numpages = {4},
  year = {2009},
  month = {Aug},
  publisher = {American Physical Society},
  doi = {10.1103/PhysRevLett.103.070401},
  url = {https://link.aps.org/doi/10.1103/PhysRevLett.103.070401}
}

@article{PhysRevB.90.081111,
  title = {Quantum phases of a chain of strongly interacting anyons},
  author = {Finch, Peter E. and Frahm, Holger and Lewerenz, Marius and Milsted, Ashley and Osborne, Tobias J.},
  journal = {Phys. Rev. B},
  volume = {90},
  issue = {8},
  pages = {081111},
  numpages = {5},
  year = {2014},
  month = {Aug},
  publisher = {American Physical Society},
  doi = {10.1103/PhysRevB.90.081111},
  url = {https://link.aps.org/doi/10.1103/PhysRevB.90.081111}
}

@article{FINCH2014299,
title = "{Integrable anyon chains: From fusion rules to face models to effective field theories}",
journal = {Nuclear Physics B},
volume = {889},
pages = {299-332},
year = {2014},
issn = {0550-3213},
doi = {https://doi.org/10.1016/j.nuclphysb.2014.10.017},
url = {https://www.sciencedirect.com/science/article/pii/S055032131400323X},
author = {Peter E. Finch and Michael Flohr and Holger Frahm},
}

@article{Zauner_2015,
doi = {10.1088/1367-2630/17/5/053002},
url = {https://dx.doi.org/10.1088/1367-2630/17/5/053002},
year = {2015},
month = {may},
publisher = {IOP Publishing},
volume = {17},
number = {5},
pages = {053002},
author = {Zauner, V and Draxler, D and Vanderstraeten, L and Degroote, M and Haegeman, J and Rams, M M and Stojevic, V and Schuch, N and Verstraete, F},
title = {Transfer matrices and excitations with matrix product states},
journal = {New J. Phys.},
}

@article{PhysRevB.55.2164,
  title = {Class of ansatz wave functions for one-dimensional spin systems and their relation to the density matrix renormalization group},
  author = {Rommer, Stefan and \"Ostlund, Stellan},
  journal = {Phys. Rev. B},
  volume = {55},
  issue = {4},
  pages = {2164--2181},
  numpages = {0},
  year = {1997},
  month = {Jan},
  publisher = {American Physical Society},
  doi = {10.1103/PhysRevB.55.2164},
  url = {https://link.aps.org/doi/10.1103/PhysRevB.55.2164}
}

@article{PhysRevB.85.100408,
  title = {Variational matrix product ansatz for dispersion relations},
  author = {Haegeman, Jutho and Pirvu, Bogdan and Weir, David J. and Cirac, J. Ignacio and Osborne, Tobias J. and Verschelde, Henri and Verstraete, Frank},
  journal = {Phys. Rev. B},
  volume = {85},
  issue = {10},
  pages = {100408},
  numpages = {5},
  year = {2012},
  month = {Mar},
  publisher = {American Physical Society},
  doi = {10.1103/PhysRevB.85.100408},
  url = {https://link.aps.org/doi/10.1103/PhysRevB.85.100408}
}

@article{PhysRevB.88.075133,
  title = "{Post-matrix product state methods: To tangent space and beyond}",
  author = {Haegeman, Jutho and Osborne, Tobias J. and Verstraete, Frank},
  journal = {Phys. Rev. B},
  volume = {88},
  issue = {7},
  pages = {075133},
  numpages = {35},
  year = {2013},
  month = {Aug},
  publisher = {American Physical Society},
  doi = {10.1103/PhysRevB.88.075133},
  url = {https://link.aps.org/doi/10.1103/PhysRevB.88.075133}
}

@article{10.21468/SciPostPhysLectNotes.7,
	title={{Tangent-space methods for uniform matrix product states}},
	author={Laurens Vanderstraeten and Jutho Haegeman and Frank Verstraete},
	journal={SciPost Phys. Lect. Notes},
	pages={7},
	year={2019},
	publisher={SciPost},
	doi={10.21468/SciPostPhysLectNotes.7},
	url={https://scipost.org/10.21468/SciPostPhysLectNotes.7},
}

@article{HOLZHEY1994443,
title = {Geometric and renormalized entropy in conformal field theory},
journal = {Nuclear Physics B},
volume = {424},
number = {3},
pages = {443-467},
year = {1994},
issn = {0550-3213},
doi = {https://doi.org/10.1016/0550-3213(94)90402-2},
url = {https://www.sciencedirect.com/science/article/pii/0550321394904022},
author = {Christoph Holzhey and Finn Larsen and Frank Wilczek},
}

@book{francesco2012conformal,
  title={Conformal field theory},
  author={Di Francesco, Philippe and Mathieu, Pierre and S{\'e}n{\'e}chal, David},
  year={2012},
  publisher={Springer New York},
  doi = {10.1007/978-1-4612-2256-9},
  url = {https://doi.org/10.1007/978-1-4612-2256-9}
}

@article{FRIEDAN198537,
title = "{Superconformal invariance in two dimensions and the tricritical Ising model}",
journal = {Physics Letters B},
volume = {151},
number = {1},
pages = {37-43},
year = {1985},
issn = {0370-2693},
doi = {https://doi.org/10.1016/0370-2693(85)90819-6},
url = {https://www.sciencedirect.com/science/article/pii/0370269385908196},
author = {Daniel Friedan and Zongan Qiu and Stephen Shenker},
}

@article{DOTSENKO198454,
title = "{Critical behaviour and associated conformal algebra of the Z3 Potts model}",
journal = {Nuclear Physics B},
volume = {235},
number = {1},
pages = {54-74},
year = {1984},
issn = {0550-3213},
doi = {https://doi.org/10.1016/0550-3213(84)90148-2},
url = {https://www.sciencedirect.com/science/article/pii/0550321384901482},
author = {Vl.S. Dotsenko},
}

@article{MPS1,
author={Fannes, M. and Nachtergaele, B. and Werner, R. F.},
title={Finitely correlated states on quantum spin chains},
journal={Commun. Math. Phys.},
year={1992},
month={Mar},
day={01},
volume={144},
number={3},
pages={443-490},
issn={1432-0916},
doi={10.1007/BF02099178},
url={https://doi.org/10.1007/BF02099178}
}

@article{MPS2,
title = {The density-matrix renormalization group in the age of matrix product states},
journal = {Annals of Physics},
volume = {326},
number = {1},
pages = {96-192},
year = {2011},
note = {January 2011 Special Issue},
issn = {0003-4916},
doi = {https://doi.org/10.1016/j.aop.2010.09.012},
url = {https://www.sciencedirect.com/science/article/pii/S0003491610001752},
author = {Ulrich Schollwöck},
}

@article{MPS3,
	title={{Efficient numerical simulations with Tensor Networks: Tensor Network Python (TeNPy)}},
	author={Johannes Hauschild and Frank Pollmann},
	journal={SciPost Phys. Lect. Notes},
	pages={5},
	year={2018},
	publisher={SciPost},
	doi={10.21468/SciPostPhysLectNotes.5},
	url={https://scipost.org/10.21468/SciPostPhysLectNotes.5},
}

@article{BoseHubbard,
  title = {Boson localization and the superfluid-insulator transition},
  author = {Fisher, Matthew P. A. and Weichman, Peter B. and Grinstein, G. and Fisher, Daniel S.},
  journal = {Phys. Rev. B},
  volume = {40},
  issue = {1},
  pages = {546--570},
  numpages = {0},
  year = {1989},
  month = {Jul},
  publisher = {American Physical Society},
  doi = {10.1103/PhysRevB.40.546},
  url = {https://link.aps.org/doi/10.1103/PhysRevB.40.546}
}

@article{BoseHubbard1,
author = {Bruder, C. and Fazio, R. and Schön, G.},
title = "{The Bose-Hubbard model: from Josephson junction arrays to optical lattices}",
journal = {Annalen der Physik},
volume = {517},
number = {9-10},
pages = {566-577},
doi = {https://doi.org/10.1002/andp.200551709-1005},
year = {2005}
}

@article{BoseHubbard2,
  title = {Superconductivity in narrow-band systems with local nonretarded attractive interactions},
  author = {Micnas, R. and Ranninger, J. and Robaszkiewicz, S.},
  journal = {Rev. Mod. Phys.},
  volume = {62},
  issue = {1},
  pages = {113--171},
  numpages = {0},
  year = {1990},
  month = {Jan},
  publisher = {American Physical Society},
  doi = {10.1103/RevModPhys.62.113},
  url = {https://link.aps.org/doi/10.1103/RevModPhys.62.113}
}

@software{MPSKit,
  author       = {Van Damme, Maarten and
                  Devos, Lukas and
                  Haegeman, Jutho},
  title        = "{MPSKit, Zenodo}",
  month        = mar,
  year         = 2025,
  publisher    = {Zenodo},
  version      = {v0.12.6},
  doi          = {10.5281/zenodo.14962291},
  url          = {https://doi.org/10.5281/zenodo.14962291},
}

\end{document}